\documentclass[11pt]{article}
\usepackage[T1]{fontenc}
\usepackage[utf8]{inputenc}
\usepackage{lmodern}
\usepackage[margin=1in]{geometry}
\usepackage{microtype,amsmath,booktabs,tabularx,longtable,graphicx}
\usepackage[numbers,sort&compress]{natbib}
\usepackage{placeins}
\usepackage{tikz}
\usetikzlibrary{arrows.meta,positioning}
\usepackage[hidelinks]{hyperref}
\usepackage{url}
\graphicspath{{figures/}}
\usepackage{authblk}

\title{A Comprehensive Survey of Crystal Structural Prototypes: Construction, Classification, and Applications}
\author[1,*]{Jianjun Hu}
\author[1]{Ying Feng}
\author[1,3]{Rongzhi Dong}
\author[2]{Jeffrey Hu}
\author[1]{Remy Rogers}
\author[1]{Dipannoy Gupta}

\affil[1]{Department of Computer Science and Engineering, University of South Carolina, Columbia, SC 29201, USA}
\affil[2]{Department of Materials Science and Engineering, University of Illinois Urbana-Champaign, Urbana, IL, USA}
\affil[3]{Current affiliation: School of Artificial Intelligence, Wuhan University of Technology, Wuhan, China}
\affil[*]{Corresponding author: \texttt{jianjunh@cse.sc.edu}}

\begin{document}
\maketitle
\begin{abstract}
Crystal structural prototypes describe recurring atomic arrangements across chemically distinct materials. Prototype libraries are constructed through crystallographic curation, structural standardization, geometric matching, local-environment clustering, and periodic-graph analysis. They support chemical substitution, derivative enumeration, high-throughput screening, diffraction-based structure solution, crystal structure prediction (CSP), and generative design. This survey compares methods by the structural information they retain and the equivalence rules they impose, distinguishing crystallographic structure types, geometric realizations, chemistry-anonymous topologies, and chemical decorations. We review curated and automatically extracted prototype libraries (AFLOW, CSPD, LAE-ICSPD, EMFDTW, TYPIX, and ICSD structure types), their source databases, and algorithms for assigning family membership. A GraphID-based analysis of samples from ICSD, the Materials Project, OQMD, and LeMat shows that the 100 most frequent families in each sample cover 30--67\% of analyzed structures, while 72--82\% of observed families are singletons. The GraphIDs of up to 76\% of structures change when the neighbor tolerance is varied. In our comparison of 12 generative CSP models, template retrieval achieves the highest top-1 success rate. Removing a reduced-stoichiometry class from training and validation introduces failures in a previously satisfied structure or symmetry criterion for 50--78\% of that class's test structures. This ablation measures sensitivity to stoichiometric coverage; it does not isolate generalization to unseen structural prototypes. We relate these findings to earlier statistics of structure types, space groups, and periodic nets, and to studies of protein-fold coverage. We propose independent-database coverage measurements, discovery-rate analyses, and structural-family-held-out tests to assess how prototype reuse affects prediction accuracy, generalization, and structural novelty.
\end{abstract}
\tableofcontents
\clearpage
\section{Introduction}
Generative artificial intelligence (AI) has become an important route to discovering new materials~\cite{Zhao2021CubicGAN,Zhao2023PGCGM}, with algorithms such as CrystalFlow~\cite{Luo2025CrystalFlow}, uFlowCSP~\cite{Dey2026uFlowCSP}, and MatterGen~\cite{Zeni2025MatterGen}, but it can propose candidates without capturing the structural principles that make crystals physically plausible~\cite{Hu2024GenerativeAI}. One such principle is the crystal structural prototype: a recurring atomic arrangement shared by chemically distinct materials, which underpins much of computational materials discovery, from high-throughput screening~\cite{PrototypeUse15} and structure solution from diffraction~\cite{PrototypeUse06} to crystal structure prediction~\cite{PrototypeUse02}. A recent study shows that modern deep generative models implicitly use structural templates when generating and predicting structures~\cite{Wei2026GenerativeCSPBenchmark}. This section illustrates the value of prototypes through familiar mineral and functional-material families, and then defines the scope of this survey and its relation to earlier reviews.
\subsection{Recurring structures in minerals and functional materials}
Prototypes connect mineralogical classification with functional materials: they describe recurring sites, coordination environments, and connectivity that help relate composition to properties. Two examples show why this recurrence is useful for materials discovery and why a family label alone does not determine performance.

Perovskites provide a versatile example. Their ABX$_3$ framework places larger A-site species within a network of corner-sharing BX$_6$ octahedra, creating distinct sites for chemical substitution and collective distortions~\cite{PrototypeCase01,PrototypeCase08}. In BaTiO$_3$, polar displacements within this framework produce ferroelectricity; across oxide and halide perovskites, changes in chemistry and geometry yield dielectric, magnetic, and optoelectronic responses. For discovery, the family turns an unrestricted arrangement problem into three narrower questions: which elements occupy each site, which octahedral tilts or off-centering patterns are favorable, and whether the resulting crystal is stable. A geometric analysis has applied this site-based reasoning to thousands of hypothetical compositions~\cite{PrototypeCase01}.

Diamond-related networks illustrate the same principle for high lattice thermal conductivity (LTC). Diamond is a reference material for high thermal conductivity, but transport also depends on bonding, atomic masses, phonon dispersion, and scattering: isotopically enriched diamond conducts heat measurably better than natural diamond with no change of structure~\cite{PrototypeCase02}. Cubic boron arsenide, whose high-quality single crystals reach about 1300~W\,m$^{-1}$\,K$^{-1}$ at room temperature~\cite{PrototypeCase03}, has the zinc-blende structure type; it differs crystallographically from elemental diamond but shares the same tetrahedral net once chemical labels are removed. This pair motivates searches across chemical decorations of a shared network, followed by chemistry-specific transport and stability calculations. Crystallographic and chemistry-anonymous classifications thus offer complementary routes to candidates.

Other families play similar roles. Garnet-type Li$_7$La$_3$Zr$_2$O$_{12}$~\cite{PrototypeCase04} and NASICON frameworks~\cite{PrototypeCase06} guide ion-conductor searches; olivine LiFePO$_4$ established a durable insertion-electrode architecture~\cite{PrototypeCase05}; spinels couple cation distribution to magnetism~\cite{Sickafus1999Spinel,PrototypeCase10}; and zeolites and metal--organic frameworks organize pore networks~\cite{PrototypeLibrary47,PrototypeUse20}. Mineral families also expose distinctions that a prototype description must preserve: feldspars accommodate substitution and Al/Si ordering~\cite{MineralFamily01}, calcite and aragonite are same-composition polymorphs, and zinc blende and wurtzite share tetrahedral coordination but differ in stacking~\cite{MineralFamily02}. Table~\ref{tab:functional-cases} summarizes these families and the information needed beyond family identity.

\begin{table}[!htbp]\centering\small
\caption{Structural families that motivate prototype-based discovery, with the information a family label does not supply.}\label{tab:functional-cases}
\begin{tabularx}{\linewidth}{p{0.21\linewidth}XX}\toprule
Family & Illustrative connection & Information beyond the family label \\ \midrule
Perovskite & BaTiO$_3$ polar response; electronic functionality~\cite{PrototypeCase01} & Chemistry, octahedral tilts, off-centering \\
Diamond / zinc-blende-related net & Diamond and BAs as high-LTC references~\cite{PrototypeCase02,PrototypeCase03} & Bonding, masses, phonon scattering, defects \\
Zinc blende / wurtzite & Shared tetrahedral coordination; different stacking~\cite{PrototypeLibrary01} & Global connectivity and geometric realization \\
Garnet-related & Mineral substitutions and LLZO lithium transport~\cite{PrototypeCase07,PrototypeCase04} & Chemical sites, mobile-ion occupancy, barriers \\
Spinel & Normal/inverse cation distribution; magnetite magnetism~\cite{Sickafus1999Spinel,PrototypeCase10} & Cation distribution, valence, magnetic order \\
Feldspar & Framework substitution and ordering variants~\cite{MineralFamily01} & Composition, Al/Si ordering, distortion \\
Calcite / aragonite & Same-composition polymorphism~\cite{MineralFamily02} & Carbonate packing and cation coordination \\
NASICON & Sodium-ion transport~\cite{PrototypeCase06} & Site populations and transport bottlenecks \\
Olivine phosphate & LiFePO$_4$ insertion electrodes~\cite{PrototypeCase05} & Redox chemistry and transport \\
Zeolite / MOF frameworks & Pore architecture and candidate assembly~\cite{PrototypeLibrary47,PrototypeUse20} & Geometry, chemistry, accessibility \\
\bottomrule\end{tabularx}\end{table}

Prototype knowledge therefore shapes both search strategies and representations. In template-based CSP, a target composition is matched to candidate families whose sites are decorated, ordered, and relaxed, using data-mined substitution probabilities~\cite{PrototypeUse22}, learned composition metrics~\cite{PrototypeUse03}, or chemical-similarity ranking~\cite{PrototypeUse02,PrototypeUse04} (Section~6.1). Generative models learn the distribution of known structures and can propose arrangements outside any explicit template list, either in coordinate space~\cite{Xie2022CDVAE,Jiao2023DiffCSP,Zeni2025MatterGen} or over symmetry and Wyckoff descriptions~\cite{PrototypeLibrary42,Jiao2024DiffCSPpp}. In inverse design, a target property can select a family---tetrahedral networks for thermal transport, garnet frameworks for ion conduction, or a net topology for gas capture~\cite{PrototypeUse21}---before chemistry and geometry are optimized. Explicit family definitions also sharpen evaluation, because they distinguish a new composition in a known arrangement from a new structural family and make family-held-out tests possible~\cite{PrototypeUse01,PrototypeUse11}.

Recurrence has long been quantified in crystal chemistry. Space groups are populated very unevenly~\cite{Urusov2009SpaceGroups}; intermetallic compounds concentrate in a few dominant structure types with a long tail of rare ones~\cite{Dshemuchadse2015Statistics}; and a handful of underlying nets account for most three-periodic coordination polymers~\cite{Alexandrov2011UnderlyingNets}. Together with the practical successes above, these statistics motivate the working hypothesis that physically relevant inorganic crystals occupy a limited effective repertoire of recurring structural families. Section~\ref{sec:repertoire} formulates the hypothesis in testable form.

\FloatBarrier
\subsection{Scope, contributions, and relation to previous reviews}
This survey treats crystal structural prototypes as reusable knowledge for materials discovery, tracing their construction, classification, and use. The primary scope is inorganic crystalline materials: mineral-derived families, crystallographic structure types, geometric similarity classes, and chemistry-anonymous periodic topologies. 
To examine whether physically relevant crystals occupy a limited repertoire of structural families, we draw methodological lessons from protein-fold studies on structural recurrence and library coverage~\cite{ProteinFold01}.

Existing CSP reviews organize the field around search methods or generative models. Oganov and colleagues survey structure-search methods and their applications, with emphasis on global optimization and quantum-mechanical prediction~\cite{CSPReview01}. De Breuck and colleagues organize crystal generative models by architecture, representation, conditioning, and materials domain, and discuss data and evaluation~\cite{CSPReview02}. Our organizing unit is the structural prototype itself, followed from mineralogical nomenclature through library curation, automated family assignment, template retrieval, derivative enumeration, and generative or inverse design. This choice connects the definition of structural identity to the behavior of the algorithms that depend on it.

The survey makes four contributions. First, it separates crystallographic types, geometric realizations, periodic topology, and chemical decoration (Section~2), and compares the algorithms that implement each equivalence rule (Table~\ref{tab:comparison-methods}). Second, it compares curated and automatically extracted libraries by source data, representative selection, retained information, and interoperability (Sections~3--4), and quantifies family statistics and definition sensitivity in four sampled collections (Section~\ref{sec:empirical-distributions}). Third, it maps how prototypes enter substitution, screening, structure solution, CSP, and generative design (Sections~5--6, Table~\ref{tab:csp-approaches}). Fourth, it turns the limited-repertoire hypothesis into measurable quantities---independent coverage, discovery rate, threshold sensitivity, and family-held-out performance---and relates them to precedents in crystal chemistry and protein structure (Sections~7--8).

\begin{center}\fbox{\begin{minipage}{0.94\linewidth}\small
\textbf{Box 1. Scope and limitations.}
\begin{itemize}\setlength{\itemsep}{1pt}
\item \emph{Review method.} This survey integrates literature on prototype construction, classification, and applications, drawing on targeted searches and citation tracing to connect developments across crystallography, materials databases, crystal structure prediction, and generative design.
\item \emph{Counts.} Prototype counts reported by different libraries use different source corpora and equivalence rules (Table~\ref{tab:libraries}); they are not estimates on a common scale.
\item \emph{Empirical analysis.} Section~\ref{sec:empirical-distributions} analyzes sampled snapshots of four collections that share upstream sources. It is descriptive: it is not a full-database census, a saturation result, or a CSP benchmark. GraphID families are fingerprint classes under stated conventions.
\item \emph{Stability and novelty.} Membership in a family says nothing about thermodynamic, dynamical, or experimental stability. Novelty is always relative to a declared reference set and comparison rule.
\item \emph{Reuse versus completeness.} Evidence that many structures reuse a few families does not establish that the set of physically realizable families is finite. The protein-fold analogy motivates testing this possibility.
\end{itemize}
\end{minipage}}\end{center}

Sections~2--4 establish definitions, construction methods, and library resources; Sections~5--6 examine applications and computational design; Section~7 addresses the limited-repertoire hypothesis and the protein analogy; and Section~8 discusses open challenges.

\section{Defining crystal prototypes}
Every library count and coverage figure in this review depends on what is taken to be a prototype, so we first set out the competing definitions. They range from crystallographic structure types, through geometric and topological notions of equivalence, to the names and families of mineralogy, and they differ in what they preserve and what they ignore. This is why later sections report each library together with its equivalence rule.

\subsection{Crystallographic structure types and equivalence rules}
The word \emph{prototype} is used for several different objects: a representative crystal, a crystallographic structure type, an equivalence class under a matching procedure, or an underlying periodic framework. The crystallographic definition is the oldest. The IUCr nomenclature report defines isopointal and isoconfigurational structures in terms of space group, occupied Wyckoff positions, and comparable geometry~\cite{PrototypeLibrary24}, and the standardization work of Parth\'e and coworkers made such comparisons operational by fixing settings and Wyckoff sequences~\cite{PrototypeLibrary22,PrototypeLibrary25}. Burzlaff and Malinovsky proposed a classification procedure built on these ideas~\cite{PrototypeLibrary26}, and the ICSD adopted structure-type assignments for its entries~\cite{PrototypeLibrary14,PrototypeLibrary12}. Anonymous stoichiometry, also used in these schemes, records ratios among distinguishable site species but not how the sites are arranged.
Every comparison therefore needs an explicit equivalence rule. Equal space groups do not imply equal structures, and different space groups can describe small distortions of one parent arrangement. A library count is interpretable only when its unit of identity and its inclusion criteria are stated.

\subsection{Geometry and symmetry}
Geometric comparison asks whether atomic sites can be mapped onto each other after cell transformations, origin shifts, and species permutations. Symmetry constrains the possible mappings, and a geometric cost measures the remaining difference. The Bilbao Crystallographic Server tools quantify the distortion between structures related by group--subgroup relations~\cite{PrototypeLibrary29}, and Thomas, Natarajan, and Van der Ven developed a general framework that enumerates symmetry-distinct mappings and scores them by strain and atomic displacement~\cite{PrototypeLibrary30}.

A single space group and Wyckoff description can still contain distinct geometric configurations. For RbCuS and RbCuSe, configurations sharing a nominal structure type differ enough that relaxing one representative misses the others~\cite{PrototypeLibrary51}. Conversely, a small distortion can change a symmetry label while preserving connectivity, which is why AFLOW-XtalFinder tests geometry rather than relying on symmetry labels alone~\cite{PrototypeLibrary06}. Crystallographic identity, geometric similarity, and topology are therefore related but distinct descriptors.

\subsection{Topology, chemical decoration, and physical state}
A topological description specifies nodes, edges, and periodic translations. Removing element identities lets chemically different realizations share a parent connectivity, as the Reticular Chemistry Structure Resource does for nets~\cite{PrototypeLibrary44} and GraphID does for bonded crystal graphs~\cite{PrototypeLibrary33}. The information discarded depends on how the graph is built: an all-atom graph and a reduced framework graph can assign the same structure to different families.

We organize structural information in four levels: topology or framework, geometric and symmetry realization, chemical decoration and ordering, and physical state (Fig.~\ref{fig:workflows}a). Existing databases implement these levels in different combinations. The Decoratypes taxonomy formalizes the decoration level, showing that structures with similar anonymous arrangements can differ in oxidation states, ordering, and magnetic configuration, and hence in electronic properties~\cite{PrototypeLibrary32}.

\subsection{Mineral names and structural classification}
Mineralogy supplies a layer of classification that complements computational catalogs. Names such as perovskite, garnet, spinel, olivine, rutile, and fluorite identify familiar arrangements, often extended from natural minerals to synthetic compounds~\cite{PrototypeLibrary24}. A mineral species is a chemically and crystallographically defined entity, whereas a mineral-derived family name covers many chemical realizations, so neither a formula nor a mineral name substitutes for a computational equivalence rule.

The IMA-approved garnet-supergroup nomenclature organizes minerals isostructural with garnet across chemical classes, distinguishing the dodecahedral, octahedral, and tetrahedral cation sites and recording substitutions systematically~\cite{PrototypeCase07}. Synthetic lithium garnets extend this language beyond silicates~\cite{PrototypeCase04}. The perovskite-supergroup classification combines structure and composition in a hierarchy that accommodates structural derivatives~\cite{PrototypeCase08}. The spinel-supergroup scheme likewise requires the characteristic polyhedral arrangement in addition to a nominal AB$_2$X$_4$ formula, so a compound with that stoichiometry but different connectivity is excluded~\cite{PrototypeCase09}. These classifications give a chemically interpretable bridge between named families and automated assignment.

For interoperable libraries, mineral family names work best as annotations linked to explicit representatives, symmetry and site descriptions, and graph identifiers. A broad named family can span several geometric or symmetry variants, so its correspondence to a topology-only graph class depends on the neighbor and framework-reduction conventions.

\subsection{Examples of prototypes from mineral families}
Named mineral families show concretely how chemistry varies within a recurring structural organization. Related members can differ in lattice parameters, distortions, ordering, or symmetry, so ``isostructural'' must refer to a declared classification level. Solid solution reflects chemically compatible site substitution, sometimes with coupled charge-compensating replacements at more than one site~\cite{PrototypeCase07,PrototypeCase09,MineralFamily01}.

Garnet and spinel organize coordination polyhedra in two different ways. The common silicate garnets pyrope, almandine, and grossular contain isolated SiO$_4$ tetrahedra linked to octahedral and eight-coordinate metal sites; the broader supergroup includes non-silicate compositions and lower-symmetry variants~\cite{PrototypeCase07}. Spinels place cations in tetrahedral and octahedral sites of a nearly cubic close-packed anion array. MgAl$_2$O$_4$, magnetite, and chromite illustrate chemical variation within this arrangement, and the normal and inverse cation distributions require separate annotation~\cite{Sickafus1999Spinel,PrototypeCase09}.

Zinc blende and wurtzite separate local coordination from global arrangement. Sphalerite ZnS and GaAs have tetrahedral coordination with cubic stacking, whereas wurtzite ZnS and greenockite CdS have hexagonal stacking~\cite{MineralFamily02}. Their underlying nets differ (\texttt{dia} versus \texttt{lon})~\cite{PrototypeLibrary44}, although both structures have tetrahedral first-neighbor coordination. The pair is a direct test of whether a prototype method preserves stacking information or groups structures by coordination alone.

Feldspars are a framework family whose decoration and ordering cannot be captured by one fixed geometry. Corner-sharing (Al,Si)O$_4$ tetrahedra form a three-dimensional framework with larger cations in the cavities. Orthoclase and microcline share the composition KAlSi$_3$O$_8$ but differ in symmetry and Al/Si ordering; amazonite is a variety of microcline. The albite--anorthite series involves the coupled substitution Na$^+$ + Si$^{4+}$ $\leftrightarrow$ Ca$^{2+}$ + Al$^{3+}$, with labradorite an intermediate composition. Temperature-dependent ordering, exsolution, and compositional substitution are distinct processes~\cite{MineralFamily01}.

Calcite and aragonite show why composition-based grouping fails. Both are CaCO$_3$, but they adopt trigonal and orthorhombic structures with different cation--carbonate arrangements. Rhodochrosite and smithsonite adopt the calcite type, whereas cerussite adopts the aragonite type~\cite{MineralFamily02}. The same-composition pair illustrates polymorphism; substitutions within either arrangement illustrate a different family relationship.

\begin{table}[htbp]\centering\footnotesize
\caption{Named mineral families and their relationship to prototypes. Symmetry refers to common representatives or parent arrangements; members of a broad family can differ in geometry and symmetry.}\label{tab:mineral-families}
\begin{tabularx}{\linewidth}{p{0.15\linewidth}p{0.16\linewidth}XX}\toprule
Family / type & Common symmetry & Structural organization & Representative members \\ \midrule
Garnet & Usually cubic; variants occur & Tetrahedral, octahedral, and eight-coordinate sites & Pyrope, almandine, grossular \\
Zinc blende & Cubic, $F\overline{4}3m$ & Tetrahedral network with cubic stacking & Sphalerite ZnS; GaAs \\
Spinel & Usually cubic parent; variants occur & Close-packed anions with tetrahedral and octahedral cation sites & MgAl$_2$O$_4$; magnetite; chromite \\
Feldspar & Monoclinic / triclinic & Corner-sharing aluminosilicate framework; chemical and ordering variants & Orthoclase; microcline; albite; anorthite \\
2H wurtzite & Hexagonal, $P6_3mc$ & Tetrahedral network with hexagonal stacking & Wurtzite ZnS; greenockite CdS \\
Calcite & Trigonal, $R\overline{3}c$ & Carbonate arrangement with calcite-type cation coordination & Calcite; rhodochrosite; smithsonite \\
Aragonite & Orthorhombic, $Pnma$ & Different carbonate packing and cation coordination & Aragonite; cerussite \\
\bottomrule\end{tabularx}
\end{table}

These examples show why a named mineral family must be linked to an explicit structural description (Fig.~\ref{fig:workflows}a). Local coordination alone misses the zinc-blende--wurtzite distinction; formula alone misses calcite--aragonite polymorphism; and a strict symmetry label splits related feldspar or garnet realizations. A useful library exposes these relationships explicitly rather than mapping every named group to a single computational class.

\begin{figure}[htbp]\centering
\begin{tikzpicture}[node distance=4mm,box/.style={draw,rounded corners,align=center,text width=4.1cm,minimum height=9mm,font=\footnotesize},lab/.style={font=\small\bfseries,align=center,text width=4.3cm}]
\node[lab] (ta) {(a) Structural information};
\node[box,below=2mm of ta] (a1) {Topology or periodic framework};
\node[box,below=of a1] (a2) {Geometry and symmetry realization};
\node[box,below=of a2] (a3) {Chemical decoration and ordering};
\node[box,below=of a3] (a4) {Physical state: pressure, temperature, magnetism};
\node[lab,right=0.5cm of ta] (tb) {(b) Library construction};
\node[box,below=2mm of tb] (b1) {Source structures and provenance};
\node[box,below=of b1] (b2) {Scope filters, standardization, duplicate audit};
\node[box,below=of b2] (b3) {Comparison: symmetry, geometry, local, invariant, or graph};
\node[box,below=of b3] (b4) {Family assignment and representatives};
\node[box,below=of b4] (b5) {Invariance and sensitivity checks; versioned release};
\node[lab,right=0.5cm of tb] (tc) {(c) Use in discovery};
\node[box,below=2mm of tc] (c1) {Templates, nets, or symmetry descriptions};
\node[box,below=of c1] (c2) {Retrieve and decorate, enumerate, or condition a generator};
\node[box,below=of c2] (c3) {Construct candidates and relax};
\node[box,below=of c3] (c4) {Reclassify; assess novelty against declared references};
\node[box,below=of c4] (c5) {Physical qualification and application tests};
\foreach \x/\y in {a1/a2,a2/a3,a3/a4,b1/b2,b2/b3,b3/b4,b4/b5,c1/c2,c2/c3,c3/c4,c4/c5} \draw[-{Stealth}] (\x)--(\y);
\end{tikzpicture}
\caption{The three views of prototypes used in this survey. (a) Levels of structural information; each level adds descriptive detail to the one above it, and the arrows do not denote a one-to-one classification or a physical transformation. (b) Construction of a prototype library; the comparison stage defines the operational family identity, and unresolved records are retained with explicit status. (c) Roles of prototypes in computational discovery; routes can be combined, relaxation can change the initial family assignment, and novelty and physical validity are assessed separately.}\label{fig:hierarchy}\label{fig:workflows}\end{figure}
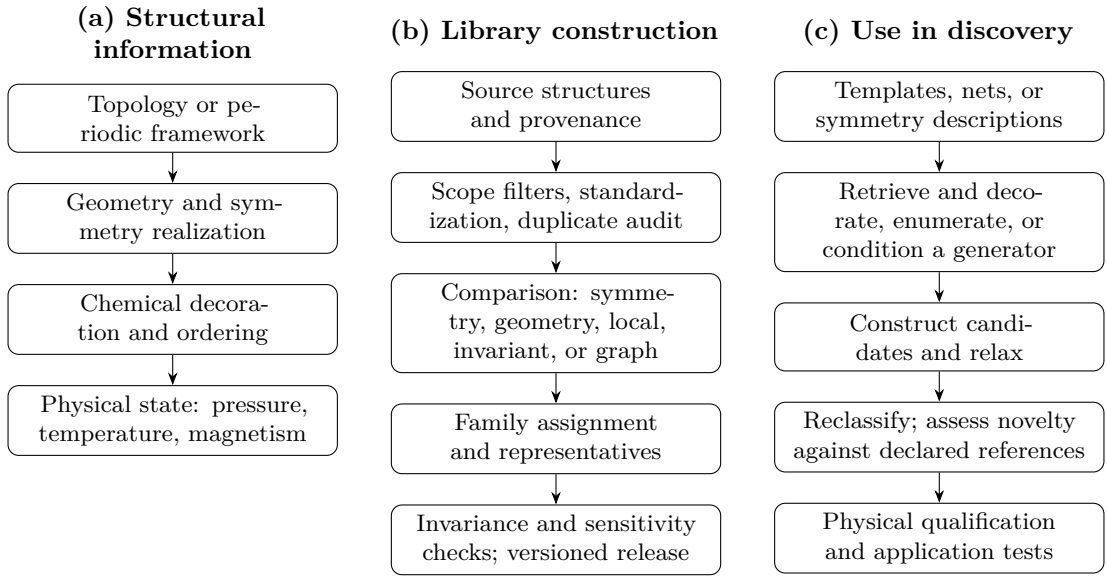
\begin{table}[htbp]\centering\small
\caption{Operational prototype definitions and their limits.}\label{tab:definitions}
\begin{tabularx}{\linewidth}{lXX}\toprule Representation & Retained information & Principal limitation \\ \midrule
Crystallographic type & Symmetry and occupied sites & One label may admit different geometries \\
Geometry cluster & Atomic mappings and distances & Tolerance-dependent families \\
Local environments & Coordination and neighborhood descriptors & Local similarity need not fix global connectivity \\
Periodic topology & Chosen nodes, edges, translations & Graph-construction and reduction dependence \\
Decorated template & Species assignment and coordinates & Chemical feasibility requires validation \\
\bottomrule\end{tabularx}\end{table}

\FloatBarrier
\begin{samepage}
\section{Building prototype libraries}

A prototype library results from a sequence of choices: which structures to include, how to standardize and deduplicate them, and which comparison rule groups them into families. After preparing the source corpus, library builders select a comparison method and choose representatives for the resulting families. The methods below compare symmetry, geometry, local environments, continuous invariants, or periodic graphs. Because each choice changes the number and meaning of the resulting prototypes, the same corpus can yield different libraries.
\par
\end{samepage}

\subsection{Source selection, standardization, and deduplication}
Crystal prototype library construction begins with a source corpus and a declared scope (Fig.~\ref{fig:workflows}b). Records of experimental and computed structures must retain their provenance, and repeated measurements or repeated imports of one compound must be identified before structural diversity is estimated. Standardization removes conventional choices---cell setting, origin, and coordinate conventions---that would otherwise make equivalent structures look different~\cite{PrototypeLibrary22,PrototypeLibrary25}. STRUCTURE TIDY implemented this standardization as a program~\cite{PrototypeLibrary23}, and modern libraries rely on symmetry finders with explicit tolerances, such as spglib~\cite{Togo2024Spglib} and AFLOW-SYM, which chooses tolerances self-consistently~\cite{Hicks2018AflowSym}.

Duplicate detection then compares standardized structures. CMPZ~\cite{PrototypeLibrary27} and XtalComp~\cite{PrototypeLibrary28} provide efficient pairwise comparison of periodic structures, and the StructureMatcher in pymatgen~\cite{Ong2013Pymatgen} is widely used in materials informatics and CSP benchmarks. Standardization does not settle physical equivalence: ordered derivatives, partial occupancies, and genuine structural changes still require separate treatment, and structures that fail parsing or matching should be reported rather than silently dropped from coverage denominators.

\subsection{Prototype extraction using symmetry and geometry}
Symmetry-led extraction uses crystallographic descriptors to restrict comparisons before global geometric matching. AFLOW-XtalFinder makes the task explicit: it separates material identity, prototype identity, decorations, and magnetic configurations, and it groups AFLOW and ICSD structures into prototypes automatically~\cite{PrototypeLibrary06}. The result is a different kind of resource from the manually curated, parameterized AFLOW encyclopedia (Section~\ref{sec:aflow}).

Geometry-led clustering captures similarities that a single symmetry label misses, at the cost of choosing a distance, a threshold, and a representative. CSPD, for example, groups COD structures by composition type and cell size before a similarity analysis~\cite{PrototypeLibrary07}. Match relations produced by pairwise matchers such as StructureMatcher need not be transitive under finite tolerances, so clustering their outputs requires an explicit linkage rule. A reproducible construction report states which comparisons are chemistry-anonymous, how site permutations are handled, and how family counts change under reasonable tolerance variations.

\subsection{Local-environment descriptors and learned clustering}
Prototype extraction methods based on local environments characterize the coordination of each site and aggregate site descriptors into a structure-level comparison. The atomic-environment-type (AET) classification of Daams and Villars assigns each site to a coordination-polyhedron class and has been used to predict the environments adopted in binary compounds~\cite{Daams2000AET,Villars2004AET}. In computational workflows, local order parameters~\cite{PrototypeLibrary31}, the CrystalNN neighbor-finding algorithm benchmarked against alternative coordination-number methods~\cite{Pan2021CrystalNN}, and the smooth overlap of atomic positions (SOAP) kernel~\cite{Bartok2013SOAP} provide interpretable or differentiable site fingerprints. LAE-ICSPD clusters such local-environment fingerprints hierarchically to extract 15,613 prototypes from ICSD and shows how the clustering threshold controls library size~\cite{PrototypeLibrary08}.

Local agreement identifies shared motifs but does not fix global periodic connectivity; zinc blende and wurtzite are the standard counterexample. A pipeline therefore needs to state whether local descriptors define families or only screen candidates before a stronger global comparison.

\subsection{Continuous invariants}
Tolerance-based structure matchers produce discontinuous family assignments: an arbitrarily small perturbation can move a structure across a threshold. Continuous isometry invariants address this directly. Average minimum distances (AMD) summarize the distances from each atom to its $k$ nearest neighbors and vary continuously under perturbation~\cite{Widdowson2022AMD}; pointwise distance distributions (PDD) retain the per-atom distributions and are generically complete for periodic point sets, which allows near-duplicate detection across large databases~\cite{Widdowson2021PDD}. These invariants support distance-based clustering, although defining discrete families still requires a threshold.

\subsection{GraphID and chemistry-anonymous graph families}
GraphID assigns identifiers to periodic crystal graphs, allowing records to be linked across large materials databases~\cite{PrototypeLibrary33}. Its topology-only option drops element labels, which makes it directly usable for chemistry-anonymous prototype construction; the default chemistry-aware identifier answers a different question and should be labeled separately.

For a topology census, the neighbor criterion, the graph dimensionality, the periodic representation, and the treatment of cell multiplicity are all part of the operational family definition. A GraphID is a fingerprint under these conventions rather than a proof of periodic-net isomorphism. An audit should test atom reordering, origin shifts, equivalent cells, supercells, and label substitutions that preserve the graph, and ambiguous cases can be checked against net-topology tools such as ToposPro~\cite{Blatov2014ToposPro} or the RCSR~\cite{PrototypeLibrary44}.

\subsection{Other comparison methods and representative selection}
EMFDTW identifies crystal structures from distance-matrix-derived functions compared by dynamic time warping and supports several comparison criteria; its 17,340 ``skeleton'' prototypes retain a reduced set of atoms rather than a complete template~\cite{PrototypeLibrary11}. Reduced representations of this kind suit retrieval but omit details needed for decoration, so the reduction belongs in any count or coverage claim.

\begin{figure}[htbp]\centering
\includegraphics[width=\linewidth,height=0.62\textheight,keepaspectratio]{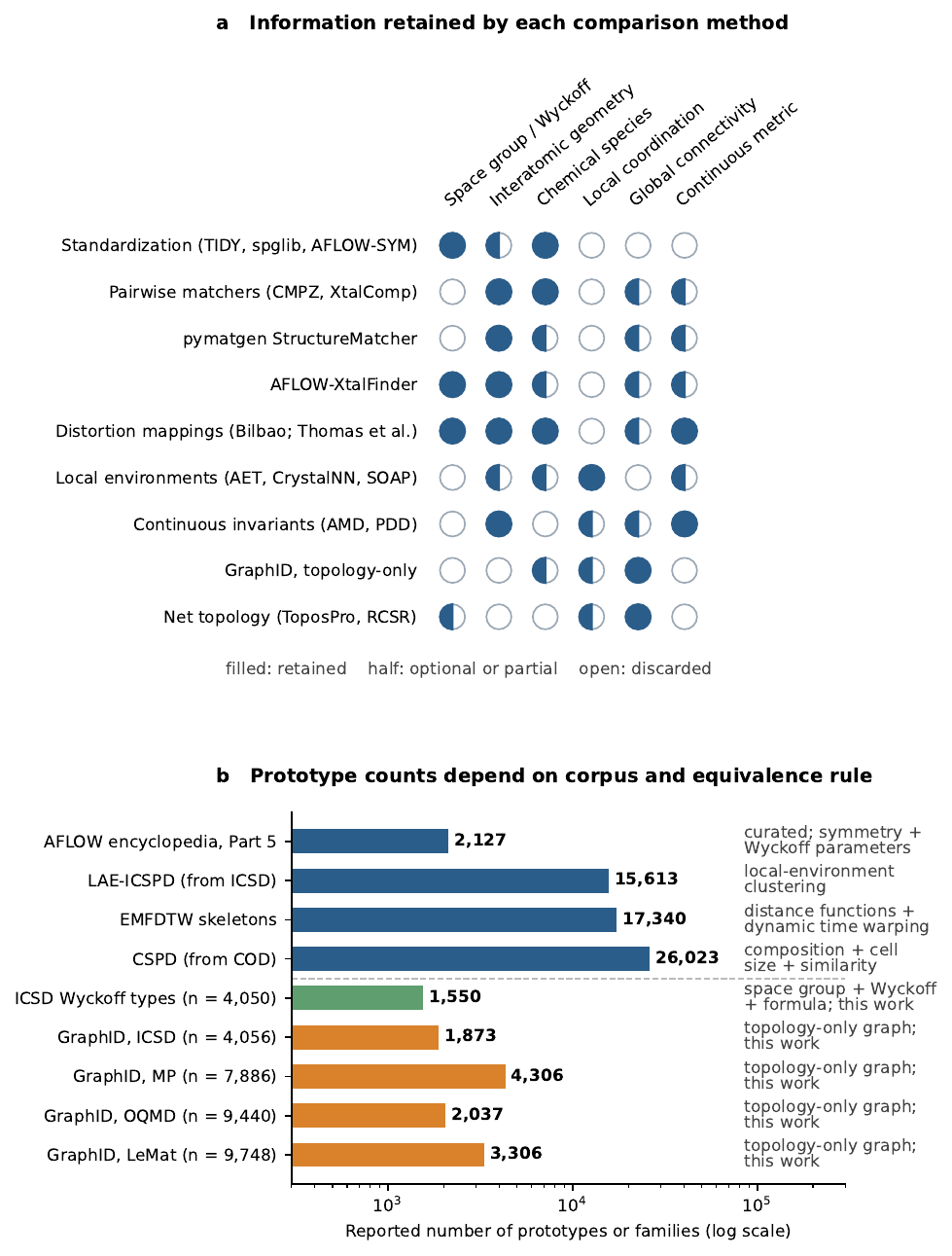}
\caption{The prototype landscape. (a) Information retained by each comparison method, summarizing Table~\ref{tab:comparison-methods}: filled circles mark information the method retains, half circles information that is optional or only partly retained, and open circles information it discards. (b) Reported prototype counts and the equivalence rules behind them. Library counts (blue) come from different corpora and rules~\cite{PrototypeLibrary05,PrototypeLibrary08,PrototypeLibrary11,PrototypeLibrary07}. Counts from this work are families observed in the samples of Section~\ref{sec:empirical-distributions}, with sample sizes $n$ given in the labels: crystallographic structure types defined by space group, Wyckoff sequence, and anonymous formula (green), and chemistry-anonymous GraphID families (orange). The crystallographic rule yields 1,550 types among 4,050 structures with successful symmetry assignments; GraphID yields 1,873 families among all 4,056 retained ICSD structures. The spread in counts reflects differences in source corpora, sample sizes, and equivalence rules.}\label{fig:landscape}
\end{figure}

Representative selection links classification to use. A representative chosen for minimal geometric redundancy may differ from one chosen for chemical compatibility, structural quality, or computational cost, and libraries can keep several representatives per family when one geometry cannot initialize all important configurations~\cite{PrototypeLibrary51}. The trade-off is between a compact catalog and faithful coverage of the configurations a task requires. Table~\ref{tab:comparison-methods} compares the main comparison methods side by side, and Fig.~\ref{fig:landscape}a condenses the comparison into the information each method retains.

{\small\setlength{\tabcolsep}{3pt}\renewcommand{\arraystretch}{1.1}
\begin{longtable}{p{0.17\linewidth}p{0.21\linewidth}p{0.17\linewidth}p{0.16\linewidth}p{0.21\linewidth}}
\caption{Structure-comparison and family-assignment methods: what each compares, what it retains, the parameters that control it, and where it fails.}\label{tab:comparison-methods}\\
\toprule
Method & Comparison basis & Retained information & Controlling parameters & Known failure mode \\ \midrule
\endfirsthead
\multicolumn{5}{l}{\textit{Table \thetable\ continued}}\\
\toprule
Method & Comparison basis & Retained information & Controlling parameters & Known failure mode \\ \midrule
\endhead
\midrule\multicolumn{5}{r}{\textit{Continued on next page}}\\\endfoot
\bottomrule\endlastfoot
Standardization (STRUCTURE TIDY; spglib; AFLOW-SYM)~\cite{PrototypeLibrary23,Togo2024Spglib,Hicks2018AflowSym} & Standard setting and Wyckoff sequence & Space group, Wyckoff letters, standardized coordinates & Symmetry tolerance & Distorted variants receive different labels; no distance between structures \\
CMPZ; XtalComp~\cite{PrototypeLibrary27,PrototypeLibrary28} & Pairwise comparison after cell reduction & Full geometry and species & Lattice and position tolerances & Duplicate decisions depend on tolerances \\
pymatgen StructureMatcher~\cite{Ong2013Pymatgen} & Lattice and site mapping after Niggli reduction; species-aware or anonymous & Geometry, optionally species & \texttt{ltol}, \texttt{stol}, \texttt{angle\_tol} (defaults 0.2, 0.3, 5$^\circ$) & Match relation need not be transitive; grouping cost grows quadratically with candidates per composition \\
AFLOW-XtalFinder~\cite{PrototypeLibrary06} & Symmetry prefilter, then geometric mapping & Prototype label, decoration, magnetic configuration & Symmetry and misfit tolerances & Symmetry tolerance can split near-identical distorted structures \\
Distortion mappings (Bilbao tools; Thomas et al.)~\cite{PrototypeLibrary29,PrototypeLibrary30} & Group--subgroup or symmetry-distinct mappings scored by strain and displacement & Distortion path and magnitude & Mapping cost function & Expensive for large cells; needs a defined parent \\
Local environments (AET; order parameters; CrystalNN; SOAP; LAE-ICSPD)~\cite{Daams2000AET,PrototypeLibrary31,Pan2021CrystalNN,Bartok2013SOAP,PrototypeLibrary08} & Per-site fingerprints aggregated and clustered & Coordination motifs & Neighbor algorithm; cluster threshold & Same first-neighbor coordination, different global connectivity (zinc blende vs.\ wurtzite) \\
Continuous invariants (AMD; PDD)~\cite{Widdowson2022AMD,Widdowson2021PDD} & Isometry invariants of interatomic distances & Geometry up to isometry & Number of neighbors $k$; family threshold & Chemistry-anonymous by default; a threshold is still needed for discrete families \\
GraphID~\cite{PrototypeLibrary33} & Identifier of periodic bonded graph & Connectivity and dimensionality; optional chemistry & Neighbor criterion and tolerance & Edge set shifts with neighbor tolerance (24.0--69.8\% identifier agreement; Section~\ref{sec:empirical-distributions}) \\
Net topology (ToposPro; RCSR)~\cite{Blatov2014ToposPro,PrototypeLibrary44} & Underlying net after simplification & Net symbol (e.g.\ \texttt{dia}, \texttt{lon}) & Simplification scheme & Discards geometry and chemistry; depends on node and edge choices \\
EMFDTW~\cite{PrototypeLibrary11} & Distance-matrix functions compared by dynamic time warping & Reduced (skeleton) geometry & Warping and similarity thresholds & Skeleton omits atoms needed for decoration \\
\end{longtable}}

\section{Existing libraries and classification resources}
Table~\ref{tab:libraries} lists the main prototype libraries and classification resources with their year, equivalence rule, reported scope, and data access. Their reported counts span more than an order of magnitude (Fig.~\ref{fig:landscape}b), because the resources differ in source corpus, scope, and equivalence rule.

{\footnotesize\setlength{\tabcolsep}{2.5pt}\renewcommand{\arraystretch}{1.1}
\begin{longtable}{p{0.17\linewidth}p{0.06\linewidth}p{0.17\linewidth}p{0.15\linewidth}p{0.1\linewidth}p{0.1\linewidth}p{0.16\linewidth}}
\caption{Prototype libraries and related classification resources. Counts are specific to each source corpus and equivalence rule. Year is that of the defining publication or first edition. Links checked October 11, 2026. n.v.: not verified from a primary source at the time of writing.}\label{tab:libraries}\label{tab:resource-directory}\\
\toprule
Resource (provider) & Year & Equivalence rule & Reported scope & Download & License & Access \\ \midrule
\endfirsthead
\multicolumn{7}{l}{\textit{Table \thetable\ continued}}\\
\toprule
Resource (provider) & Year & Equivalence rule & Reported scope & Download & License & Access \\ \midrule
\endhead
\midrule\multicolumn{7}{r}{\textit{Continued on next page}}\\\endfoot
\bottomrule\endlastfoot
AFLOW Prototype Encyclopedia~\cite{PrototypeLibrary01,PrototypeLibrary04,PrototypeLibrary05} (AFLOW consortium, Duke; NRL origin) & 2017--2026 & Curated; space group and Wyckoff positions with free parameters & 2,127 prototypes (Part 5) & CIF and POSCAR per prototype & n.v. & \url{https://aflow.org/p/} \\
AFLOW-XtalFinder~\cite{PrototypeLibrary06} (AFLOW consortium, Duke) & 2021 & Symmetry prefilter, then geometric misfit; optional decoration and magnetism & Automatic grouping of AFLOW and ICSD structures & Software in AFLOW & Open-source AFLOW code & \url{https://aflow.org/} \\
CSPD~\cite{PrototypeLibrary07} (Jilin Univ.; Beijing CSRC) & 2017 & Composition type and cell size, then geometric similarity & 26,023 prototypes from COD & n.v. & n.v. & \url{https://doi.org/10.1088/1361-648X/aa63cd} \\
LAE-ICSPD~\cite{PrototypeLibrary08} (Jilin Univ.) & 2022 & Local-environment fingerprints, hierarchical clustering & 15,613 prototypes from ICSD & Supporting data & n.v. & \url{https://doi.org/10.1021/acs.jpca.2c03416} \\
JAMIP / integrated ICSPD~\cite{PrototypeLibrary09} (Jilin Univ.) & 2021 & Inherits LAE-ICSPD & Generation and analysis infrastructure & Code repository & n.v. & \url{https://github.com/kzhoujlu/JAMIP-manual} \\
EMFDTW skeleton prototypes~\cite{PrototypeLibrary11} (Shanghai Univ.) & 2024 & Distance-matrix functions compared by dynamic time warping & 17,340 skeleton prototypes & Supporting data & n.v. & \url{https://doi.org/10.1021/acs.cgd.4c00346} \\
CAMD~\cite{PrototypeLibrary10} (Toyota Research Institute) & 2020 & Symmetry- and Wyckoff-based candidate generation from prototypes & Candidate library inside an autonomous discovery loop & Open-source code & n.v. & \url{https://doi.org/10.1039/D0SC01101K} \\
ICSD structure types~\cite{PrototypeLibrary12,PrototypeLibrary13} (FIZ Karlsruhe) & 2007 & Assigned structure type: space group, Wyckoff sequence, stoichiometry & Curated experimental structures & Subscription & Proprietary & \url{https://icsd.products.fiz-karlsruhe.de/} \\
TYPIX~\cite{PrototypeLibrary15,PrototypeLibrary16} (Univ.\ Geneva; Springer/Gmelin) & 1993--1994 & Standardized structure types & Printed reference, four volumes & No (book) & Publisher copyright & \url{https://doi.org/10.1007/978-3-662-02909-1} \\
Pauling File~\cite{PrototypeLibrary17,PrototypeLibrary18,Villars2019PaulingFile} (MPDS; NIMS) & 1998 & Editor-assigned structure prototypes & Inorganic structures, phase diagrams, properties & Subscription & Proprietary & \url{https://paulingfile.com/} \\
Pearson's Crystal Data~\cite{PrototypeLibrary20} (ASM Int.; MPDS; Crystal Impact) & 1991 & Assigned structure types & Inorganic structural records & Licensed software & Proprietary & \url{https://crystalimpact.com/pcd/} \\
RCSR~\cite{PrototypeLibrary44} (Arizona State; ANU; UC Berkeley) & 2008 & Net topology (symbol per net) & Curated periodic nets & Web access & n.v. & \url{https://rcsr.net/} \\
ToposPro / TTD collection~\cite{Blatov2014ToposPro} (Samara Center for Theoretical Materials Science) & 2014 & Underlying net after simplification & Taxonomy of nets in real structures & Software & n.v. & \url{https://doi.org/10.1021/cg500498k} \\
EPINET~\cite{PrototypeLibrary45} (ANU) & 2009 & Nets enumerated from hyperbolic tilings & Enumerated frameworks & Web access & n.v. & \url{https://epinet.anu.edu.au/} \\
SACADA~\cite{PrototypeLibrary46} (Hoffmann, Kabanov, Golov, Proserpio) & 2016 & Net topology of carbon allotropes & Reported carbon allotropes & n.v. & n.v. & \url{https://doi.org/10.1002/anie.201600655} \\
Deem zeolite-like database~\cite{PrototypeLibrary47,PrototypeLibrary48} (Rice Univ.) & 2009--2011 & Enumerated tetrahedral frameworks, energy-screened & Hypothetical zeolite frameworks & n.v. & n.v. & \url{https://doi.org/10.1039/C0CP02255A} \\
C2DB~\cite{PrototypeLibrary49,PrototypeLibrary50} (DTU) & 2018--2021 & 2D structures grouped by prototype and composition & 2D structures with calculated properties & Database download & n.v. & \url{https://c2db.fysik.dtu.dk/} \\
\end{longtable}}

\subsection{Historical catalogs and crystal-chemical reference works}
Historical crystallographic catalogs established the practice of organizing compounds by recurring arrangements. Strukturbericht symbols, whose history Mehl reviews~\cite{PrototypeLibrary21}, Pearson's handbooks~\cite{PrototypeLibrary20}, TYPIX~\cite{PrototypeLibrary15,PrototypeLibrary16}, the Landolt--B\"ornstein structure-type volumes~\cite{PrototypeLibrary19}, and the Pauling File~\cite{PrototypeLibrary17,PrototypeLibrary18,Villars2019PaulingFile} provide complementary nomenclature, standardized structural data, and crystal-chemical context. They record relationships among compounds and the terminology needed to recognize prior structural knowledge, not only coordinates.

These resources are best compared by purpose rather than by raw entry counts: a handbook record, a standardized structure type, and a downloadable computational representative are different units. Historical classification also supports provenance, because a modern candidate can correspond to a previously described arrangement even when its composition or database label is new~\cite{PrototypeLibrary24}.

\subsection{The AFLOW prototype encyclopedia}\label{sec:aflow}
The AFLOW Library of Crystallographic Prototypes gives curated, parameterized prototypes with symbolic lattice parameters and atomic coordinates. It grew from 288 representatives in Part~1~\cite{PrototypeLibrary01} to cumulative totals of 590, 1,100, and 1,783 in Parts~2--4~\cite{PrototypeLibrary02,PrototypeLibrary03,PrototypeLibrary04}, and Part~5 brings the total to 2,127~\cite{PrototypeLibrary05}. Part~2 completed the representation of all 230 space groups~\cite{PrototypeLibrary02}.

The encyclopedia is most useful when a workflow needs explicit representatives with consistent labels. It is distinct from the much larger AFLOW repository~\cite{Curtarolo2012AFLOW,Curtarolo2012AFLOWLIB} and from the prototype sets that AFLOW-XtalFinder extracts automatically~\cite{PrototypeLibrary06}.

\subsection{CSPD, LAE-ICSPD, and discovery platforms}
CSPD constructs prototypes from filtered COD structures by composition-type and cell-size grouping followed by similarity analysis, yielding 26,023 prototypes~\cite{PrototypeLibrary07}. LAE-ICSPD instead clusters local environments of filtered ICSD structures into 15,613 prototypes and documents the trade-off between template reduction and clustering fidelity~\cite{PrototypeLibrary08}. Because their corpora and equivalence rules differ, the two counts do not provide a direct comparison of structural coverage.

JAMIP connects these prototype resources to high-throughput generation and analysis~\cite{PrototypeLibrary09}, and CAMD builds a reusable symmetry- and Wyckoff-based candidate library inside an autonomous discovery loop~\cite{PrototypeLibrary10}. Library construction and use often happen in the same platform, so lineage and shared source structures need to be recorded when such resources are compared.

\subsection{Source databases}
Prototype libraries are extracted from source databases whose own construction shapes the families they contain. The ICSD is the principal curated collection of experimental inorganic structures~\cite{Hellenbrandt2004ICSD,PrototypeLibrary13}; its introduction of structure types connected classification directly to an established source~\cite{PrototypeLibrary12}. The Crystallography Open Database (COD) provides an open-access alternative and is the corpus behind CSPD~\cite{Grazulis2009COD,Grazulis2012COD,PrototypeLibrary07}.

Computed databases add DFT-relaxed structures and hypothetical compounds. The Materials Project relaxes structures largely drawn from the ICSD~\cite{DatasetMP}. OQMD contains ICSD-derived structures together with hypothetical compounds generated by decorating commonly occurring prototypes~\cite{DatasetOQMD,Kirklin2015OQMD}. The AFLOW framework and its AFLOWLIB repository follow a similar high-throughput strategy~\cite{Curtarolo2012AFLOW,Curtarolo2012AFLOWLIB}. Alexandria extends known prototypes through machine-learning-assisted element substitution followed by DFT validation~\cite{Schmidt2023Alexandria,Schmidt2024Alexandria}, GNoME combines symmetry-aware substitutions with random structure search~\cite{PrototypeUse08}, Materials Cloud hosts curated computational datasets~\cite{Talirz2020MaterialsCloud}, and LeMat-Bulk harmonizes several of these sources into one dataset~\cite{DatasetLeMat}. Because many computed entries are themselves decorations of known prototypes, family statistics measured on computed databases partly reflect the prototypes used to generate them. The concentration of families in Section~\ref{sec:empirical-distributions} is consistent with this construction history.

\subsection{Periodic nets and specialized libraries}
Specialized resources show especially clearly that the appropriate structural representation depends on the domain. The RCSR classifies periodic nets~\cite{PrototypeLibrary44}, EPINET enumerates further nets from hyperbolic tilings~\cite{PrototypeLibrary45}, and ToposPro determines the underlying nets of real structures~\cite{Blatov2014ToposPro}; topological surveys of the Cambridge Structural Database with these tools show that a few nets dominate coordination polymers~\cite{Alexandrov2011UnderlyingNets}. SACADA catalogs carbon allotropes and uses topology to flag structures that were reported before under different names~\cite{PrototypeLibrary46}. Net labels omit chemical and geometric information that complete crystal templates retain.

Hypothetical zeolite databases expand framework possibilities by computational enumeration~\cite{PrototypeLibrary47,PrototypeLibrary48}, whereas C2DB organizes two-dimensional materials and their calculated properties~\cite{PrototypeLibrary49,PrototypeLibrary50}. The resources differ in dimensionality, chemical scope, and evidence of physical realization: a hypothetical net, a relaxed candidate, and an experimentally characterized material are different evidence classes even when they share a prototype label.

\subsection{Prototype distributions in four database samples}\label{sec:empirical-distributions}
To illustrate the statistical questions behind prototype reuse, we compared chemistry-anonymous graph families in our downloaded local snapshots of four collections: 151,707 ICSD files, 210,579 Materials Project files, 1,076,926 OQMD files, and 3,126,657 files in the PBE directory of LeMat-Bulk~\cite{Hellenbrandt2004ICSD,DatasetMP,DatasetOQMD,DatasetLeMat}. These are counts of local files that we downloaded, not current database sizes or deduplicated materials. Fixed-seed reservoir sampling selected 10,000 files from each collection. The OQMD files are the OQMD-derived entries of the all-functional LeMat-Bulk CIF dump, identified by their source identifiers, rather than a separate OQMD download. Because LeMat aggregates MP, OQMD, and Alexandria, the four cohorts are not independent.

All cohorts were processed identically. We retained ordered structures with one to four elements and at most 52 sites in the supplied cell, erased chemical labels, and reduced to the primitive cell at a tolerance of 0.01~\AA. GraphID~\cite{PrototypeLibrary33} was then applied with geometry-based nearest-neighbor detection, a 10~\AA\ cutoff, and a primary relative neighbor tolerance of 0.10, with tolerances of 0.05 and 0.20 as sensitivity checks; the identifier includes the maximum dimensionality of connected components. Invariance was checked on ten retained structures per source: all 160 tests of atom reordering, origin shift, supercell construction, and species replacement returned identical identifiers.

Table~\ref{tab:distribution-cohorts} gives the retained denominators and observed family counts. ICSD lost 4,304 structures to the disorder filter, 145 to parsing failures, and 1,495 to the element and site limits; MP, OQMD, and LeMat lost 2,114, 560, and 252 structures to the scope limits. The disorder filter removes 43\% of the ICSD sample, so the ICSD results describe its ordered subset.

\begin{table}[htbp]\centering\small

\caption{Cohorts used for the prototype-distribution analysis. Family counts use the primary GraphID neighbor tolerance of 0.10.}\label{tab:distribution-cohorts}
\begin{tabularx}{\linewidth}{Xrrr}\toprule
Source collection & Selected files & Retained structures & Observed families \\ \midrule
ICSD & 10,000 & 4,056 & 1,873 \\
MP & 10,000 & 7,886 & 4,306 \\
OQMD (LeMat-Bulk entries) & 10,000 & 9,440 & 2,037 \\
LeMat (PBE directory) & 10,000 & 9,748 & 3,306 \\
\bottomrule\end{tabularx}\end{table}

\begin{figure}[htbp]\centering
\includegraphics[width=\linewidth]{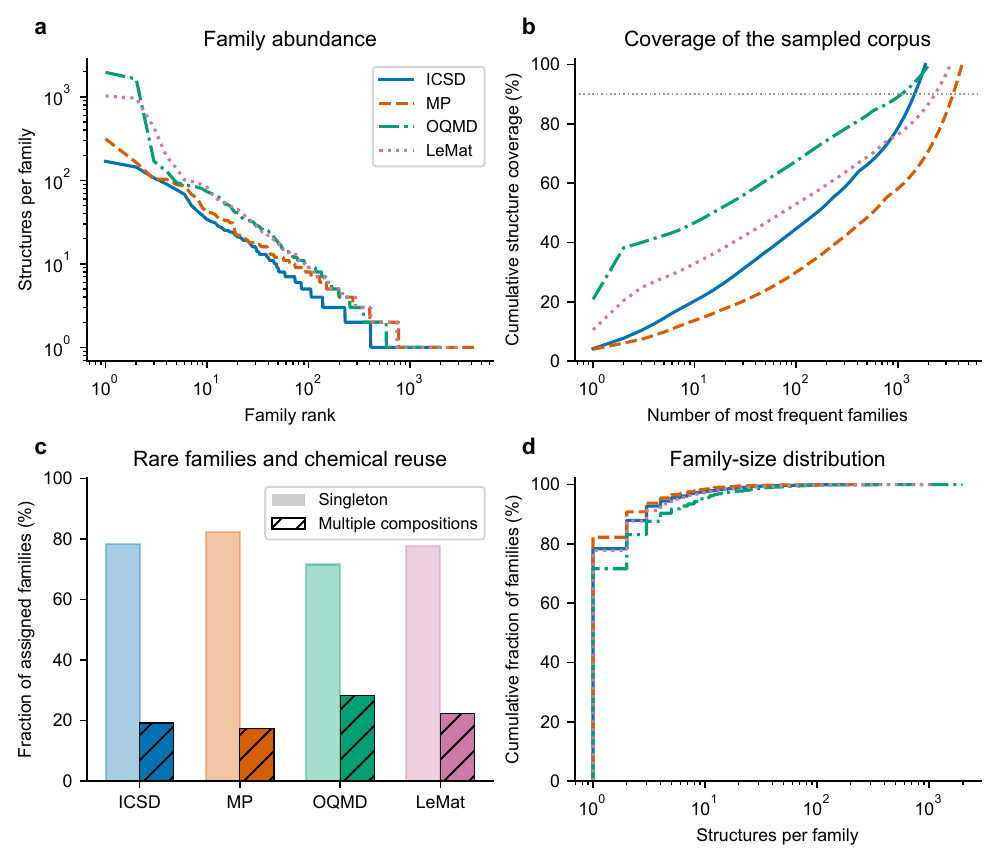}
\caption{Prototype-family abundance in the retained cohorts. (a) Structures per family ranked by abundance. (b) Cumulative structure coverage by the most frequent families; the dotted line marks 90\% coverage. (c) Fractions of singleton families and of families containing multiple distinct reduced formulas; the two categories are not an exhaustive partition. (d) Empirical cumulative family-size distributions. All panels use the primary chemistry-anonymous GraphID definition; cohorts are described in Table~\ref{tab:distribution-cohorts}.}\label{fig:prototype-abundance}
\end{figure}

Figure~\ref{fig:prototype-abundance} shows records concentrated in common families together with long rare-family tails. The 100 most frequent families cover 44.6\% of the retained ICSD sample, 30.0\% of MP, 67.4\% of OQMD, and 52.9\% of LeMat, while singletons make up 78.3\%, 82.2\%, 71.6\%, and 77.7\% of observed families, respectively. The rank--abundance curves in Fig.~\ref{fig:prototype-abundance}(a) are heavy-tailed, qualitatively matching the structure-type statistics reported for intermetallic compounds~\cite{Dshemuchadse2015Statistics}. For library design, frequent families cover a large share of records, while rare families account for most distinct families. The stronger concentration in OQMD is expected from its construction: many of its hypothetical entries are decorations of a fixed set of common prototypes~\cite{Kirklin2015OQMD}.

\begin{figure}[htbp]\centering
\includegraphics[width=\linewidth]{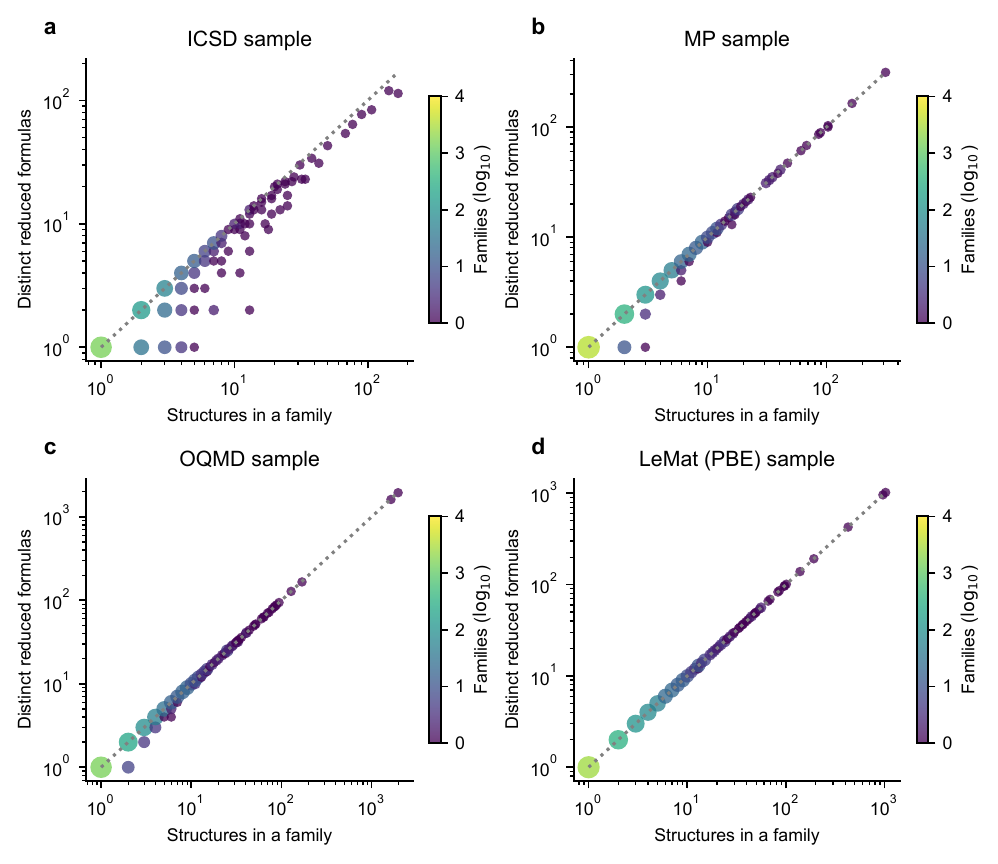}
\caption{Chemical diversity within operational prototype families in the ICSD, MP, OQMD, and LeMat samples. Each point aggregates families sharing the same pair of counts: structural records and distinct reduced formulas. Point area and color encode the number of families at that count pair on a common logarithmic scale. The dotted diagonal represents one distinct formula per record.}\label{fig:prototype-chemistry}
\end{figure}

Figure~\ref{fig:prototype-chemistry} connects family recurrence to chemical diversification. Multiple distinct reduced formulas occur in 359 ICSD families, 754 MP families, 573 OQMD families, and 738 LeMat families. Points near the diagonal are families in which almost every record has a different composition; points below it reflect repeated formulas, which in ICSD include repeated determinations of the same compound. Many recurring anonymous arrangements contain multiple compositions, although repeated records of the same composition also contribute to family abundance. The presence of multiple compositions within a family supports the use of decoration and retrieval in discovery.

\begin{figure}[htbp]\centering
\includegraphics[width=\linewidth]{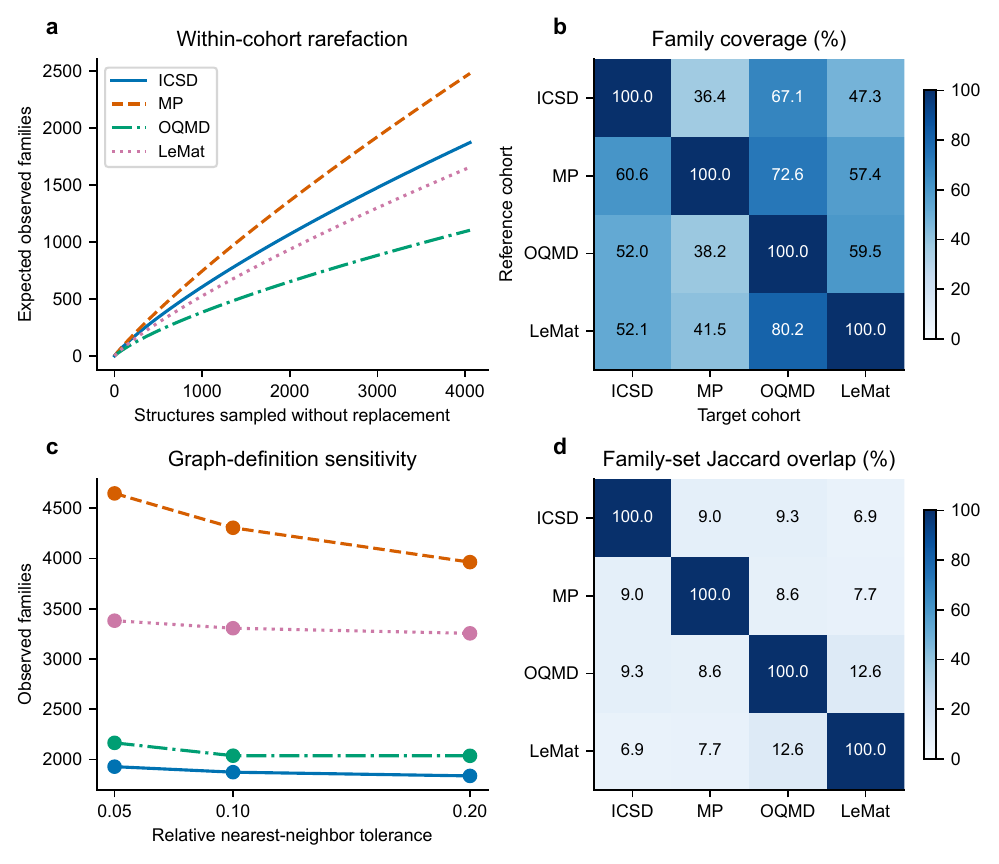}
\caption{Prototype accumulation, cross-cohort coverage, and definition sensitivity. (a) Exact expected within-cohort rarefaction under sampling without replacement over a common range of 1--4,056 records. (b) Directional coverage: each cell gives the percentage of target-cohort structures whose primary family occurs in the reference cohort. (c) Observed family counts under three relative nearest-neighbor tolerances. (d) Family-set Jaccard overlap (intersection over union). LeMat shares upstream sources with MP and OQMD, and the OQMD cohort is drawn from the LeMat-Bulk dump.}\label{fig:prototype-coverage}
\end{figure}

Within-cohort rarefaction (Fig.~\ref{fig:prototype-coverage}a) controls for the number of classified records by computing the expected number of families under sampling without replacement, up to the smallest cohort of 4,056 records. It does not control for chemistry, cell size, or lineage. The shallow OQMD curve again reflects prototype-based generation within that sample. Cross-cohort comparisons separate structure-weighted coverage from family-set overlap: families observed in the ICSD sample cover 67.1\% of OQMD records, yet the family-set Jaccard overlap of the two cohorts is only 9.3\%. Common families dominate record coverage even when the rare-family sets barely overlap.

As a crystallographic reference, we also grouped the retained ICSD structures by the classical structure-type key on which ICSD assignments build---space group, Wyckoff sequence, and anonymous formula~\cite{PrototypeLibrary24,PrototypeLibrary12}---computed with spglib~\cite{Togo2024Spglib} in the standardized setting. 
The 4,050 structures with a symmetry result (six failed) form 1,550 Wyckoff-sequence types at a symmetry tolerance of 0.01~\AA\ and 1,544 at 0.1~\AA, compared with 1,873 GraphID families among all 4,056 retained ICSD structures (Fig.~\ref{fig:landscape}b). Among the 4,050 structures with symmetry assignments, 68\% of Wyckoff-sequence types are singletons and the 100 most frequent types cover 47\% of structures. The corresponding GraphID statistics for the full 4,056-structure cohort are 78\% and 44.6\%, respectively. The most frequent types are familiar ones: rock salt, ThCr$_2$Si$_2$, MgCu$_2$ Laves, CsCl, Cu$_3$Au, PbFCl, GdFeO$_3$-type perovskite, fluorite, ZrNiAl, and cubic perovskite. The two partitions largely agree (adjusted Rand index 0.71, V-measure 0.94). GraphID families are nearly pure with respect to Wyckoff type (homogeneity 0.95), and Wyckoff types split across GraphID families slightly more often (completeness 0.93), consistent with GraphID's larger family count. Because the Wyckoff letters are not normalized over equivalent origin choices, a few equivalent types may be split, so 1,550 is an upper bound for this rule.

Family counts depend on the graph definition (Fig.~\ref{fig:prototype-coverage}c). Moving the neighbor tolerance from 0.05 to 0.20 gives 1,928--1,836 ICSD families, 4,648--3,964 MP families, 2,165--2,036 OQMD families, and 3,381--3,255 LeMat families. Aggregate counts change modestly, but individual identifiers change substantially: agreement with the primary setting is 42.7--69.8\% at tolerance 0.05 and 24.0--41.3\% at 0.20. Partition-based comparisons are needed to determine how often structures move between families.

\FloatBarrier

\section{Using prototypes for materials discovery}
Once a library has been constructed, its prototypes can generate candidate structures for new compositions. Discovery workflows use them to substitute elements, enumerate ordered variants, screen candidates, and interpret diffraction data. Each application depends on the structural information retained by the prototype definition.

\subsection{Chemical substitution and candidate construction}
Chemical substitution converts a known atomic arrangement into a candidate for a new composition. Hautier and coworkers learned the probability of ionic substitutions from known compounds~\cite{PrototypeUse22} and combined such a model with DFT to predict previously unreported ternary oxides~\cite{PrototypeUse23}. The approach exploits structural recurrence while leaving a chemical selection problem: charge balance, coordination preferences, atomic sizes, and electronic effects decide whether a decoration is viable, which is why template methods add chemical ranking and relaxation steps~\cite{PrototypeUse02,PrototypeUse03}. Combinatorial screening of all stoichiometric compositions shows how large the chemical space is before any structure is assigned~\cite{Davies2016Screening}.

\subsection{Derivative structures, ordering, and occupancy}
Derivative-structure algorithms enumerate the symmetrically distinct decorations of a parent lattice. Hart and Forcade introduced the approach for simple lattices~\cite{PrototypeLibrary35} and extended it to multilattices~\cite{PrototypeLibrary36} and fixed concentrations~\cite{PrototypeLibrary37}; symmetry reduction removes equivalent configurations and makes the candidate spaces tractable. Enumeration is complete only within the declared parent lattice, cell size, species, and occupancy constraints. Relaxation can turn a derivative ordering into a different geometric or connectivity realization, so both the parent family and the relaxed classification should be recorded~\cite{PrototypeLibrary32}. Disorder classification provides the complementary view from partially occupied experimental structures~\cite{PrototypeLibrary34}.

\subsection{Screening, active learning, and property-directed discovery}
Prototype pools supply structured candidate sets for property-directed searches, and they enter workflows in different ways. Flores and coworkers built IrO$_2$ and IrO$_3$ candidates from prototype structures and used active learning to find stable polymorphs for oxygen evolution~\cite{PrototypeUse15}. CAMD generates candidates from a symmetry- and Wyckoff-based library inside an autonomous loop~\cite{PrototypeLibrary10}. Vasylenko and coworkers instead extract element features from local geometries in known crystals~\cite{PrototypeUse16}, and Li and coworkers screen organic--inorganic hybrid heterostructures within selected structural families~\cite{PrototypeUse17}. For high-entropy carbides and ceramics, entropy descriptors computed over decorations of a fixed rocksalt-type parent rank compositions for synthesis~\cite{PrototypeUse18,PrototypeUse19}. Prototypes thus serve as candidate generators, feature sources, or fixed parents for composition screening, and a workflow description should state which role they play.

\subsection{Structure solution from diffraction}
Prototype-assisted structure solution uses known arrangements to constrain the interpretation of diffraction data. Griesemer, Ward, and Wolverton matched powder patterns against prototype-derived candidates in a high-throughput solution workflow~\cite{PrototypeUse06}. The prior is powerful when the target belongs to an available family and restrictive when it does not. Parackal and coworkers address the second case by combining symmetry-compatible Wyckoff enumeration, machine-learning ranking, coordinate fitting, and DFT to solve structures beyond known prototypes~\cite{PrototypeUse07}.

\subsection{Topology-guided framework assembly and inverse design}
For metal--organic frameworks, periodic nets organize building-block assembly: Col\'on, G\'omez-Gualdr\'on, and Snurr construct frameworks automatically from net topologies and evaluate them for energy applications~\cite{PrototypeUse20}. In inverse design, the topology becomes a discrete choice alongside metal clusters and linkers, as in reinforcement-learning design of frameworks for direct air capture of CO$_2$~\cite{PrototypeUse21}. The design variable is a reduced framework rather than a full crystal template, so linker geometry, interpenetration, flexibility, and chemical compatibility decide whether a chosen net yields the intended material.

\section{Prototypes in crystal structure prediction and generative modeling}
In crystal structure prediction, prototypes can provide templates for chemical decoration and baselines for evaluating generative models. Template methods retrieve known arrangements, whereas generative methods learn structural distributions from data. Both require a declared reference library and comparison rule to assess structural novelty.

\subsection{Template retrieval, chemical ranking, and relaxation}
Template-based CSP separates retrieval of plausible arrangements from optimization of their chemical realizations. Our TCSP method retrieves chemically compatible templates from the Materials Project and ranks them by chemical similarity and oxidation-state compatibility~\cite{PrototypeUse02,Fu2023OxidationState}. TCSP~2.0 extends the template pool to Materials Cloud, C2DB, and GNoME structures and adds improved oxidation-state prediction, element embeddings, and chemical ranking~\cite{PrototypeUse04}; its reported improvements reflect changes to both the template pool and candidate ranking. CSPML learns a metric over composition pairs to choose substitution templates before local relaxation~\cite{PrototypeUse03}, and PNcsp+ selects candidates through periodic-number similarity enhanced by machine learning~\cite{PrototypeUse05}. A retrieved structure is only an initial candidate, and relaxation can preserve or change its connectivity. Template availability, retrieval quality, and post-relaxation agreement are therefore separate quantities: a large library cannot help if retrieval ranks the right family low, and a good ranking cannot recover a family absent from the pool. Figure~\ref{fig:workflows}c summarizes these roles, and Table~\ref{tab:csp-approaches} compares how prototype knowledge enters the main CSP and generation approaches.

\begin{table}[htbp]\centering\small
\caption{How prototype knowledge enters CSP and generative approaches, and at which structural level each can produce something new.}\label{tab:csp-approaches}
\begin{tabularx}{\linewidth}{p{0.18\linewidth}XXp{0.22\linewidth}}\toprule
Approach & Where prototype knowledge enters & Level at which novelty is possible & Representative methods \\ \midrule
Template retrieval and substitution & Explicit template library ranked by chemical similarity & Composition and decoration in a known arrangement & TCSP, TCSP 2.0, CSPML, PNcsp+, ionic substitution~\cite{PrototypeUse02,PrototypeUse04,PrototypeUse03,PrototypeUse05,PrototypeUse22} \\
Derivative enumeration & Parent lattice or prototype & Ordering and decoration within the parent & Hart--Forcade enumeration~\cite{PrototypeLibrary35,PrototypeLibrary36,PrototypeLibrary37} \\
Composition generator plus templates & Template library supplies structures for generated formulas & Composition and decoration within the template repertoire & 2D generative pipelines~\cite{PrototypeUse12,PrototypeUse13} \\
Symmetry- and Wyckoff-based generation & Learned distribution over space groups and Wyckoff occupations & New combinations of symmetry, occupation, and geometry & Wyckoff Transformer, WyCryst, DiffCSP++, symmetry-aware diffusion, hierarchical GFlowNets~\cite{PrototypeLibrary42,PrototypeLibrary43,Jiao2024DiffCSPpp,PrototypeUse10,PrototypeUse09} \\
Topology and protostructure enumeration & Nets or protostructures enumerated combinatorially & New connectivity within the enumerated space & Topology-based generation; 39-billion protostructure screen~\cite{PrototypeLibrary39,PrototypeLibrary41} \\
Coordinate-space generative models & Implicit, through the training distribution & Any level in principle & CDVAE, DiffCSP, FlowMM, MatterGen, CrystaLLM~\cite{Xie2022CDVAE,Jiao2023DiffCSP,Miller2024FlowMM,Zeni2025MatterGen,Antunes2024CrystaLLM} \\
Global structure search & None required; physics-based energy landscape & Any level & Evolutionary and random search~\cite{CSPReview01} \\
\bottomrule\end{tabularx}\end{table}

\subsection{Composition generators coupled to structural templates}
A crystal composition generator can propose candidate compositions~\cite{Wei2024CCT,Fu2023MaterialTransformers}, which can be fed to a separate template method, such as TCSP~\cite{PrototypeUse02} or CSPML~\cite{PrototypeUse03}, to predict their structures. Dong and coworkers generate 2D-material compositions with a transformer and assign structures by template-based prediction~\cite{PrototypeUse12}; Lyngby and Thygesen combine deep generative models with lattice decoration of known 2D prototypes~\cite{PrototypeUse13}; and Shen, Su, and He analyze the structure prototypes of computed two-dimensional ferromagnets~\cite{PrototypeUse14}. In such pipelines the structural repertoire is inherited from the template library, so a new formula is not a new arrangement. Reports should state whether novelty concerns composition, decoration, geometry, or topology.

\subsection{Crystal generation using symmetry, topology, and atomic coordinates}
Coordinate-free descriptions represent candidates by space group and occupied Wyckoff sites before explicit coordinates exist. Jain and Bligaard's atomic-position-independent descriptor~\cite{PrototypeLibrary38} and Wren~\cite{PrototypeLibrary40} use these descriptions for learning and coarse-grained stability screening. Generative models then sample them directly: the Wyckoff Transformer~\cite{PrototypeLibrary42} and WyCryst~\cite{PrototypeLibrary43} generate Wyckoff representations, DiffCSP++ constrains diffusion to a chosen space group and Wyckoff template~\cite{Jiao2024DiffCSPpp}, Ishii and coworkers condition diffusion on symmetry~\cite{PrototypeUse10}, and hierarchical generative flow networks make sequential decisions over space group, lattice, and atoms~\cite{PrototypeUse09}.

Topology-based generation builds structures from periodic nets and symmetry operations~\cite{PrototypeLibrary39}, and protostructure enumeration explores bounded combinatorial spaces at scale: Parackal and coworkers screened 39 billion protostructures with machine-learned stability models~\cite{PrototypeLibrary41}. Enumerated protostructures are candidate descriptions, so their counts measure the size of a search space rather than the number of realizable families.

Coordinate-space generative models learn lattices and atomic positions; some also generate species, whereas others condition on composition. CDVAE introduced diffusion-based decoding of periodic materials~\cite{Xie2022CDVAE}; DiffCSP applies joint equivariant diffusion to lattices and fractional coordinates for structure prediction at fixed composition~\cite{Jiao2023DiffCSP}; FlowMM replaces diffusion with Riemannian flow matching~\cite{Miller2024FlowMM}; MatterGen adds property conditioning and reports the rate of stable, unique, and novel structures against a combined reference set~\cite{Zeni2025MatterGen}; and CrystaLLM generates CIF text autoregressively~\cite{Antunes2024CrystaLLM}. These models have no explicit template list, but their benchmarks (Perov-5, Carbon-24, MP-20, and the chronological MPTS-52 split~\cite{Baird2024MPTS}) are drawn from the same databases as prototype libraries. Whether their successes reflect structural generalization or recall of training prototypes is therefore a question that prototype-aware evaluation can answer (Section~\ref{sec:baselines}).

\subsection{Baselines and tests of structural generalization}\label{sec:baselines}
Generative discovery needs to be compared with strong reuse baselines under the same candidate and validation budgets. Szymanski and Bartel showed that decorating AFLOW prototypes and data-driven ion exchange are competitive with generative models at producing stable materials, although the stable structures they produce mostly resemble known compounds~\cite{PrototypeUse01}. Matbench Discovery evaluates machine-learned stability predictors on a test set generated largely by element substitution into known structures, so its scores measure performance within a prototype-derived distribution~\cite{Riebesell2025Matbench}. GNoME reported hundreds of thousands of new stable crystals~\cite{PrototypeUse08}, and subsequent analyses argued that many predicted or claimed new compounds are ordered versions of known disordered phases or close variants of known prototypes~\cite{Cheetham2024GNoME,Leeman2024Challenges}. Negishi and Walsh formalize the distinction by separating training duplicates, substitution-derived structures, and unmatched candidates~\cite{PrototypeUse11}.

Our benchmark compares 12 generative CSP models with TCSP~2.0 and examines sensitivity to the removal of reduced-stoichiometry classes from training~\cite{Wei2026GenerativeCSPBenchmark}. Under identical structure-matching, symmetry, and consensus criteria, template retrieval was the strongest single method (68.3\% top-1 success), ahead of the symmetry-aware EquiCSP~\cite{Lin2024EquiCSP} (66.4\%) and Uni-3DAR~\cite{Lu2025Uni3DAR} (62.9\%). Most structures solved by a generative model were also solved by template substitution. To measure the effect of withholding a reduced-stoichiometry class, we removed all training and validation structures in one class (AB$_2$, AB$_3$, ABC$_2$, or ABC$_4$) from MP-20 and retrained EquiCSP, the strongest generative model in this benchmark, from scratch. On test structures from the removed class, the retrained model achieved 33--58\% success by StructureMatcher and 22--50\% when the space group also had to match. For 50--78\% of each class's test structures, ablation changed at least one previously successful criterion (structure matching or space-group agreement) to failure (Table~\ref{tab:ablation}). These percentages describe newly introduced failures, not the net decrease in either success rate. These results show sensitivity to the training distribution, but do not by themselves establish whether the models can generate unseen structural prototypes. Structural-family-held-out tests and matched-size removal controls would address that question.

\begin{table}[htbp]\centering\footnotesize\setlength{\tabcolsep}{4pt}
\caption{Reduced-stoichiometry-class ablation in EquiCSP~\cite{Wei2026GenerativeCSPBenchmark}. The model was retrained on MP-20 after all training and validation structures in one reduced-stoichiometry class were removed, then evaluated on test structures from that class. \emph{Fit}: StructureMatcher success; \emph{consensus}: structure and space group both match; \emph{degraded}: test structures on which ablation changes at least one previously successful criterion (structure matching or space-group agreement) to failure. All percentages use the total number of test structures in that class as the denominator.}\label{tab:ablation}
\begin{tabular}{llrrrrr}\toprule
Removed class & Reduced stoichiometry & Train removed & Test & Ablated fit & Ablated consensus & Degraded \\ \midrule
ABC$_2$ & 1:1:2 ternary & 4,017 & 12 & 7/12 (58\%) & 6/12 (50\%) & 6/12 (50\%) \\
ABC$_4$ & 1:1:4 ternary & 820 & 7 & 4/7 (57\%) & 3/7 (43\%) & 4/7 (57\%) \\
AB$_3$ & 1:3 binary & 1,614 & 9 & 3/9 (33\%) & 2/9 (22\%) & 7/9 (78\%) \\
AB$_2$ & 1:2 binary & 1,393 & 17 & 7/17 (41\%) & 6/17 (35\%) & 10/17 (59\%) \\
\bottomrule\end{tabular}\end{table}

To distinguish structural generalization from reuse, test sets should separate new compositions in known families, entirely held-out families, and extrapolation in chemistry or cell size. Structural families should be excluded from model-training data and from the retrieval library before evaluation. Matched-size random-removal controls should distinguish the effect of removing a family from that of reducing the amount of training data. Computational budget, relaxation protocol, validity failures, and exposure through pretrained models should be reported alongside success rates.

\section{Is the effective crystal prototype repertoire limited?}\label{sec:repertoire}
An open question for library design and crystal discovery is whether the number of relevant structural families is effectively bounded. We separate recurrence, coverage of an independent corpus, and finiteness of the space of realizable families, and then look to protein structural biology, which faced the same question for folds. Recurrence is well supported, coverage can be tested only for a declared corpus and equivalence rule, and finiteness follows from neither.

\subsection{Structural recurrence and library coverage}
Three propositions are often conflated: that many records reuse a small number of families; that a bounded catalog covers a declared fraction of an independent corpus; and that the space of physically realizable families is finite. The first is well established, both in the curated-library literature~\cite{PrototypeLibrary01,PrototypeLibrary07,PrototypeLibrary08} and in earlier crystal-chemical statistics: space groups are populated very unevenly~\cite{Urusov2009SpaceGroups}, intermetallic structure types follow a heavy-tailed frequency distribution~\cite{Dshemuchadse2015Statistics}, a few underlying nets dominate coordination polymers~\cite{Alexandrov2011UnderlyingNets}, and the Pauling File project tracks structure-prototype statistics across its curated inorganic data~\cite{Villars2019PaulingFile}. Pauling's rule of parsimony concerns the number of distinct constituents within a crystal~\cite{Pauling1929Principles}; its relevance to the number of families across crystals is indirect. The second proposition requires independent corpora and has rarely been measured, and the third cannot follow from the first two.

Two quantities make the second proposition testable. Let $C(K)$ be the fraction of an independent target corpus assigned to $K$ families selected from a reference corpus, and let $F(N)$ be the number of distinct families observed after $N$ eligible structures, with $F(t)$ its accumulation over a defined chronology. Section~\ref{sec:empirical-distributions} reports within-cohort rarefaction and cross-cohort coverage, the sample-based analogues of $F(N)$ and $C(K)$. Network analysis of materials discovery dates~\cite{Aykol2019Network} shows that a chronological treatment is feasible. Interpreting either quantity requires controlling for chemistry, cell complexity, thermodynamic conditions, sampling, and the family definition.

\subsection{Lessons from protein folds and the structural families}
Protein structural biology asked the same question and offers tested methods for answering it. Chothia estimated about one thousand protein families~\cite{ProteinFold01}, later work extrapolated fold counts from the growth of solved structures~\cite{ProteinFold02,ProteinFold03}, and Zhang and Skolnick showed that the PDB already contained close templates for most single-domain proteins~\cite{ProteinFold04,ProteinFold05}. Analyses of AlphaFold predictions quantified both coverage and novelty. Bordin and colleagues assigned 92\% of 369,512 domains from 21 model organisms to existing CATH superfamilies~\cite{ProteinAI01}, while large-scale clustering and searches for new folds found residual novelty alongside this broad reuse~\cite{ProteinAI06,ProteinAI07}. Earlier work showed that whether fold space looks discrete or continuous depends on the comparison threshold~\cite{ProteinFold07}. Three lessons carry over to crystals: coverage must be measured on held-out data at a stated resolution; fold-count estimators that extrapolate from accumulation curves can be applied to $F(N)$ and $F(t)$; and the separation between backbone design and sequence design~\cite{ProteinAI04,ProteinAI05} mirrors the separation between structural template and chemical decoration in Section~6. Periodicity, diverse bonding, and an open chemical and cell-size space mean that the conclusions themselves do not transfer.

\section{Challenges and future directions}
This section discusses three limitations of prototype libraries. The first is bias in growth and coverage estimates arising from database construction, chronology, and the rare-family tail. The second concerns structural features that a static ordered template cannot represent, such as disorder and state dependence. The third is the infrastructure needed for interoperable libraries and reproducible benchmarks.

\subsection{Sampling bias, chronology, and the discovery of rare families}
An apparent slowdown in new-family accumulation can arise from repeated sampling of abundant families, narrow chemistry, or a database generated by decorating a fixed pool, as the OQMD sample in Section~\ref{sec:empirical-distributions} illustrates. Discovery-rate analyses should therefore report growth as a function of sampled structures alongside calendar-time curves, balance comparisons by chemistry and complexity, and check nominally independent sources for shared records and lineages~\cite{PrototypeUse11,DatasetLeMat}.

MPTS-52, a chronological split of Materials Project structures used to benchmark generative models~\cite{Baird2024MPTS,Jiao2023DiffCSP}, is a natural test bed for a temporal analysis. In our audit of its metadata, the 40,476 entries carry earliest associated reference years from 1913 to 2013. That year is a bibliographic proxy for first structural report rather than a verified discovery date. A GraphID analysis of family accumulation over these years, with random-order controls and independent post-2013 additions, would provide a chronological estimate of $F(t)$ for the specified inorganic corpus and family definition.

\subsection{Disorder, state dependence, and complex structures}
An ordered static template cannot represent partial occupancy, configurational disorder, or a fluctuating finite-temperature structure. Antypov and coworkers classify structural disorder in experimental structures directly~\cite{PrototypeLibrary34}, and high-entropy materials make the relation between a parent arrangement and its configurational realizations central~\cite{PrototypeUse18,PrototypeUse19}. Disorder also matters for novelty: ordered computational predictions that correspond to known disordered phases have been counted as new materials~\cite{Leeman2024Challenges,Cheetham2024GNoME}.

Pressure, temperature, magnetic state, and cell complexity also change the relevant repertoire. A library can cover ambient ordered compounds well while omitting high-pressure phases or long-period ordering. Coverage statements should therefore be scoped explicitly, parent families should be linked to their state-dependent realizations, and thermodynamic stability, dynamical stability, and experimental evidence of synthesis or persistence should be reported separately.

\subsection{Library interoperability and reproducible benchmarks}
An interoperable crystal prototype resource should retain its source structures, equivalence settings, representative-selection rule, and lineage across versions~\cite{PrototypeLibrary06,PrototypeLibrary33}. Linking topology identifiers to crystallographic and chemical descriptions would support both broad family grouping and application-specific distinctions, and continuous invariants~\cite{Widdowson2021PDD} could support comparisons that change smoothly under small structural perturbations. Updates should distinguish genuinely new arrangements from revised representatives or changed comparison rules.

Future work should measure coverage on independent corpora, quantify threshold sensitivity at the level of family membership, inspect unmatched candidates, and test prediction when whole families are withheld~\cite{PrototypeUse01,PrototypeUse11}. Generative and global-search methods can test the coverage of a reference library by proposing structures outside it. These candidates require separate checks of structural novelty and physical validity.

\section{Conclusions}
Crystal prototypes connect structural classification with materials discovery. Curated representatives, geometry-based clusters, local-environment families, continuous invariants, and periodic graphs retain different structural information and support different tasks, including comparison, decoration, prediction, and design (Tables~\ref{tab:definitions} and~\ref{tab:comparison-methods}).

Substitution, enumeration, and template retrieval remain useful discovery methods and strong baselines for generative models~\cite{PrototypeUse01,Wei2026GenerativeCSPBenchmark}. Our sampled collections show frequent reuse of common families alongside many rare families, consistent with earlier structure-type statistics~\cite{Dshemuchadse2015Statistics}. These observations do not establish that the space of physically realizable families is finite, and they depend on the classification rule. Across the tested changes in neighbor tolerance, GraphID identifiers change for 30.2--76.0\% of retained structures. Partition-based comparisons are needed to determine whether those changes alter family membership. The stoichiometric ablation in Section~\ref{sec:baselines} shows sensitivity to training coverage, but does not isolate generalization to unseen structural families. Independent and chronological coverage measurements, explicit equivalence rules, and structural-family-held-out benchmarks with matched-size removal controls would make the limited-repertoire hypothesis testable within a declared materials domain.

\section*{Data availability}
The curated resource list that accompanies this survey, together with the library table (Table~\ref{tab:libraries}) in machine-readable form, the bibliography, and the family-level results of Section~\ref{sec:empirical-distributions}, is available at \url{https://github.com/usccolumbia/awesome-crystal-prototype}. The results comprise per-collection statistics, the table of observed GraphID families with record counts, family-set overlap and coverage matrices, and rarefaction curves. The structure files analyzed in Section~\ref{sec:empirical-distributions} are not redistributed: the ICSD is a licensed database, and the Materials Project, OQMD, and LeMat-Bulk are available from their providers under their own terms.

\section{Contribution}
Conceptualization, J.H.; methodology, J.H., Y.F., R.D., JF.H. ; software, Y.F., J.H.; resources, J.H.; writing--original draft preparation, J.H., Y.F., R.R., JF.H., D.G.; writing--review and editing, J.H., R.R., Y.F., D.G.; visualization, J.H., Y.F.; supervision, J.H.;  funding acquisition, J.H.

\section{Declaration of conflict of interests}
None

\section*{Acknowledgement}
The research reported in this work was partially supported by the National Science Foundation under grants 2311202 and 2320292. The views, perspectives and content do not necessarily represent the official views of the NSF.The authors gratefully acknowledge the computational resources provided by the Theia high performance computing cluster at the University of South Carolina which is supported by National Science Foundation Major Research Instrumentation Grant No. 2320292. We also acknowledge the technical assistance and resources provided by Research Computing at the University of South Carolina (RRID:SCR\_027488).

\bibliographystyle{unsrtnat}
\bibliography{manuscript_references}

@article{PrototypeLibrary01,
  author = {Michael J. Mehl and David Hicks and Cormac Toher and Ohad Levy and Robert M. Hanson and Gus L. W. Hart and Stefano Curtarolo},
  title = {{The AFLOW Library of Crystallographic Prototypes: Part 1}},
  journal = {Computational Materials Science},
  volume = {136},
  pages = {S1--S828},
  year = {2017},
  doi = {10.1016/j.commatsci.2017.01.017}
}

@article{PrototypeLibrary02,
  author = {David Hicks and Michael J. Mehl and Eric Gossett and Cormac Toher and Ohad Levy and Robert M. Hanson and Gus L. W. Hart and Stefano Curtarolo},
  title = {{The AFLOW Library of Crystallographic Prototypes: Part 2}},
  journal = {Computational Materials Science},
  volume = {161},
  pages = {S1--S1011},
  year = {2019},
  doi = {10.1016/j.commatsci.2018.10.043}
}

@article{PrototypeLibrary03,
  author = {David Hicks and Michael J. Mehl and Marco Esters and Corey Oses and Ohad Levy and Gus L. W. Hart and Cormac Toher and Stefano Curtarolo},
  title = {{The AFLOW Library of Crystallographic Prototypes: Part 3}},
  journal = {Computational Materials Science},
  volume = {199},
  pages = {110450},
  year = {2021},
  doi = {10.1016/j.commatsci.2021.110450}
}

@article{PrototypeLibrary04,
  author = {Hagen Eckert and Simon Divilov and Michael J. Mehl and David Hicks and Adam C. Zettel and Marco Esters and Xiomara Campilongo and Stefano Curtarolo},
  title = {{The AFLOW Library of Crystallographic Prototypes: Part 4}},
  journal = {Computational Materials Science},
  volume = {240},
  pages = {112988},
  year = {2024},
  doi = {10.1016/j.commatsci.2024.112988}
}

@article{PrototypeLibrary05,
  author = {Nicholas H. Anderson and Michael J. Mehl and Hagen Eckert and Simon Divilov and Xiomara Campilongo and Stefano Curtarolo},
  title = {{The AFLOW Library of Crystallographic Prototypes: Part 5}},
  journal = {Computational Materials Science},
  volume = {275},
  pages = {115022},
  year = {2026},
  doi = {10.1016/j.commatsci.2026.115022}
}

@article{PrototypeLibrary06,
  author = {David Hicks and Cormac Toher and Denise C. Ford and Frisco Rose and Carlo De Santo and Ohad Levy and Michael J. Mehl and Stefano Curtarolo},
  title = {{AFLOW-XtalFinder: a reliable choice to identify crystalline prototypes}},
  journal = {npj Computational Materials},
  volume = {7},
  number = {1},
  pages = {30},
  year = {2021},
  doi = {10.1038/s41524-020-00483-4}
}

@article{PrototypeLibrary07,
  author = {Chuanxun Su and Jian Lv and Quan Li and Hui Wang and Lijun Zhang and Yanchao Wang and Yanming Ma},
  title = {{Construction of crystal structure prototype database: methods and applications}},
  journal = {Journal of Physics: Condensed Matter},
  volume = {29},
  number = {16},
  pages = {165901},
  year = {2017},
  doi = {10.1088/1361-648X/aa63cd}
}

@article{PrototypeLibrary08,
  author = {Shulin Luo and Bangyu Xing and Muhammad Faizan and Jiahao Xie and Kun Zhou and Ruoting Zhao and Tianshu Li and Xinjiang Wang and Yuhao Fu and Xin He and Jian Lv and Lijun Zhang},
  title = {{Inorganic Crystal Structure Prototype Database Based on Unsupervised Learning of Local Atomic Environments}},
  journal = {The Journal of Physical Chemistry A},
  volume = {126},
  number = {26},
  pages = {4300--4312},
  year = {2022},
  doi = {10.1021/acs.jpca.2c03416}
}

@article{PrototypeLibrary09,
  author = {Xin-Gang Zhao and Kun Zhou and Bangyu Xing and Ruoting Zhao and Shulin Luo and Tianshu Li and Yuanhui Sun and Guangren Na and Jiahao Xie and Xiaoyu Yang and Xinjiang Wang and Xiaoyu Wang and Xin He and Jian Lv and Yuhao Fu and Lijun Zhang},
  title = {{JAMIP: an artificial-intelligence aided data-driven infrastructure for computational materials informatics}},
  journal = {Science Bulletin},
  volume = {66},
  number = {19},
  pages = {1973--1985},
  year = {2021},
  doi = {10.1016/j.scib.2021.06.011}
}

@article{PrototypeLibrary10,
  author = {Joseph H. Montoya and Kirsten T. Winther and Raul A. Flores and Thomas Bligaard and Jens S. Hummelshøj and M. Aykol},
  title = {{Autonomous intelligent agents for accelerated materials discovery}},
  journal = {Chemical Science},
  volume = {11},
  number = {32},
  pages = {8517--8532},
  year = {2020},
  doi = {10.1039/D0SC01101K}
}

@article{PrototypeLibrary11,
  author = {Bing He and Yao Meng and Zhuming Gong and Kaixuan Wang and Zhou Jiang and Maxim Avdeev and Siqi Shi},
  title = {{EMFDTW: An Automated Crystallographic Identification Tool Supporting Multiple Comparison Criteria}},
  journal = {Crystal Growth \& Design},
  volume = {24},
  number = {13},
  pages = {5559--5568},
  year = {2024},
  doi = {10.1021/acs.cgd.4c00346}
}

@article{PrototypeLibrary12,
  author = {Rudolf Allmann and Roland Hinek},
  title = {{The introduction of structure types into the Inorganic Crystal Structure Database ICSD}},
  journal = {Acta Crystallographica Section A: Foundations of Crystallography},
  volume = {63},
  number = {5},
  pages = {412--417},
  year = {2007},
  doi = {10.1107/S0108767307038081}
}

@article{PrototypeLibrary13,
  author = {D. Zagorac and H. Müller and S. Ruehl and J. Zagorac and S. Rehme},
  title = {{Recent developments in the Inorganic Crystal Structure Database: theoretical crystal structure data and related features}},
  journal = {Journal of Applied Crystallography},
  volume = {52},
  number = {5},
  pages = {918--925},
  year = {2019},
  doi = {10.1107/S160057671900997X}
}

@article{PrototypeLibrary14,
  author = {G. Bergerhoff and M. Berndt and K. Brandenburg and T. Degen},
  title = {{Concerning inorganic crystal structure types}},
  journal = {Acta Crystallographica Section B: Structural Science},
  volume = {55},
  number = {2},
  pages = {147--156},
  year = {1999},
  doi = {10.1107/S0108768198010969}
}

@article{PrototypeLibrary15,
  author = {Ekkehard Fluck},
  title = {{Inorganic Crystal Structure Database (ICSD) and Standardized Data and Crystal Chemical Characterization of Inorganic Structure Types (TYPIX)---Two Tools for Inorganic Chemists and Crystallographers}},
  journal = {Journal of Research of the National Institute of Standards and Technology},
  volume = {101},
  number = {3},
  pages = {217},
  year = {1996},
  doi = {10.6028/jres.101.022}
}

@book{PrototypeLibrary16,
  author = {Erwin Parthé and Louise Gelato and Bernard Chabot and Marinella Penzo and Karin Cenzual and Roman Gladyshevskii},
  title = {{TYPIX---Standardized Data and Crystal Chemical Characterization of Inorganic Structure Types}},
  series = {Gmelin Handbook of Inorganic and Organometallic Chemistry},
  publisher = {Springer-Verlag},
  address = {Berlin, Heidelberg},
  year = {1993},
  doi = {10.1007/978-3-662-02909-1},
  note = {Four volumes, 1993--1994}
}

@article{PrototypeLibrary17,
  author = {P. Villars and N. Onodera and S. Iwata},
  title = {{The Linus Pauling file (LPF) and its application to materials design}},
  journal = {Journal of Alloys and Compounds},
  volume = {279},
  number = {1},
  pages = {1--7},
  year = {1998},
  doi = {10.1016/S0925-8388(98)00605-7}
}

@article{PrototypeLibrary18,
  author = {P. Villars and M. Berndt and K. Brandenburg and K. Cenzual and J. Daams and F. Hulliger and T. Massalski and H. Okamoto and K. Osaki and A. Prince and H. Putz and S. Iwata},
  title = {{The Pauling File, Binaries Edition}},
  journal = {Journal of Alloys and Compounds},
  volume = {367},
  number = {1-2},
  pages = {293--297},
  year = {2004},
  doi = {10.1016/j.jallcom.2003.08.058}
}

@book{PrototypeLibrary19,
  editor = {P. Villars and K. Cenzual},
  title = {{Structure Types. Part 1: Space Groups (230) $Ia\bar{3}d$--(219) $F\bar{4}3c$}},
  series = {Landolt-B{\"o}rnstein -- Group III Condensed Matter},
  volume = {43A1},
  publisher = {Springer-Verlag},
  address = {Berlin, Heidelberg},
  year = {2004},
  doi = {10.1007/b11232}
}

@book{PrototypeLibrary20,
  author = {P. Villars and L. D. Calvert},
  title = {{Pearson’s Handbook of Crystallographic Data for Intermetallic Phases}},
  edition = {Second},
  publisher = {ASM International},
  address = {Materials Park, OH},
  year = {1991},
  note = {Four volumes}
}

@article{PrototypeLibrary21,
  author = {Michael J. Mehl},
  title = {{A Brief History of Strukturbericht Symbols and Other Crystallographic Classification Schemes}},
  journal = {Journal of Physics: Conference Series},
  volume = {1290},
  number = {1},
  pages = {012016},
  year = {2019},
  doi = {10.1088/1742-6596/1290/1/012016}
}

@article{PrototypeLibrary22,
  author = {E. Parthé and L. M. Gelato},
  title = {{The standardization of inorganic crystal-structure data}},
  journal = {Acta Crystallographica Section A: Foundations of Crystallography},
  volume = {40},
  number = {3},
  pages = {169--183},
  year = {1984},
  doi = {10.1107/S0108767384000416}
}

@article{PrototypeLibrary23,
  author = {L. M. Gelato and E. Parthé},
  title = {{STRUCTURE TIDY---a computer program to standardize crystal structure data}},
  journal = {Journal of Applied Crystallography},
  volume = {20},
  number = {2},
  pages = {139--143},
  year = {1987},
  doi = {10.1107/S0021889887086965}
}

@article{PrototypeLibrary24,
  author = {J. Lima-de-Faria and E. Hellner and F. Liebau and E. Makovicky and E. Parthé},
  title = {{Nomenclature of inorganic structure types. Report of the International Union of Crystallography Commission on Crystallographic Nomenclature Subcommittee on the Nomenclature of Inorganic Structure Types}},
  journal = {Acta Crystallographica Section A: Foundations of Crystallography},
  volume = {46},
  number = {1},
  pages = {1--11},
  year = {1990},
  doi = {10.1107/S0108767389008834}
}

@article{PrototypeLibrary25,
  author = {E. Parthé and K. Cenzual and R. E. Gladyshevskii},
  title = {{Standardization of crystal structure data as an aid to the classification of crystal structure types}},
  journal = {Journal of Alloys and Compounds},
  volume = {197},
  number = {2},
  pages = {291--301},
  year = {1993},
  doi = {10.1016/0925-8388(93)90049-S}
}

@article{PrototypeLibrary26,
  author = {H. Burzlaff and Y. Malinovsky},
  title = {{A Procedure for the Clasification of Non-Organic Crystal Structures. I. Theoretical Background}},
  journal = {Acta Crystallographica Section A: Foundations of Crystallography},
  volume = {53},
  number = {2},
  pages = {217--224},
  year = {1997},
  doi = {10.1107/S0108767396013852}
}

@article{PrototypeLibrary27,
  author = {R. Hundt and J. C. Schön and M. Jansen},
  title = {{CMPZ---an algorithm for the efficient comparison of periodic structures}},
  journal = {Journal of Applied Crystallography},
  volume = {39},
  number = {1},
  pages = {6--16},
  year = {2006},
  doi = {10.1107/S0021889805032450}
}

@article{PrototypeLibrary28,
  author = {David C. Lonie and Eva Zurek},
  title = {{Identifying duplicate crystal structures: XtalComp, an open-source solution}},
  journal = {Computer Physics Communications},
  volume = {183},
  number = {3},
  pages = {690--697},
  year = {2012},
  doi = {10.1016/j.cpc.2011.11.007}
}

@article{PrototypeLibrary29,
  author = {Gemma de la Flor and Danel Orobengoa and Emre Tasci and Juan Manuel Perez-Mato and Mois I. Aroyo},
  title = {{Comparison of structures applying the tools available at the Bilbao Crystallographic Server}},
  journal = {Journal of Applied Crystallography},
  volume = {49},
  number = {2},
  pages = {653--664},
  year = {2016},
  doi = {10.1107/S1600576716002569}
}

@article{PrototypeLibrary30,
  author = {John C. Thomas and Anirudh Raju Natarajan and Anton Van der Ven},
  title = {{Comparing crystal structures with symmetry and geometry}},
  journal = {npj Computational Materials},
  volume = {7},
  number = {1},
  pages = {164},
  year = {2021},
  doi = {10.1038/s41524-021-00627-0}
}

@article{PrototypeLibrary31,
  author = {Nils E. R. Zimmermann and Anubhav Jain},
  title = {{Local structure order parameters and site fingerprints for quantification of coordination environment and crystal structure similarity}},
  journal = {RSC Advances},
  volume = {10},
  number = {10},
  pages = {6063--6081},
  year = {2020},
  doi = {10.1039/C9RA07755C}
}

@article{PrototypeLibrary32,
  author = {Kyle D. Miller and Michele Campbell and Danilo Puggioni and James M. Rondinelli},
  title = {{Decoratypes: An extensible crystal taxonomy for machine learning-guided materials discovery}},
  journal = {Physical Review Materials},
  volume = {9},
  number = {10},
  pages = {104404},
  year = {2025},
  doi = {10.1103/14yr-q3z1}
}

@article{PrototypeLibrary33,
  author = {Koki Muraoka and Taku Tanimoto and Tsubasa Munekata and Akira Nakayama},
  title = {{Universal graph-based identifiers of chemical structures for linking large material databases}},
  journal = {Nature Communications},
  volume = {17},
  number = {1},
  pages = {5409},
  year = {2026},
  doi = {10.1038/s41467-026-74536-5}
}

@article{PrototypeLibrary34,
  author = {Dmytro Antypov and Chris M. Collins and Matthew S. Dyer and John B. Claridge and Matthew J. Rosseinsky},
  title = {{Classification and statistical analysis of structural disorder in crystalline materials}},
  journal = {Journal of Applied Crystallography},
  volume = {58},
  number = {3},
  pages = {659--677},
  year = {2025},
  doi = {10.1107/S1600576725003000}
}

@article{PrototypeLibrary35,
  author = {Gus L. W. Hart and Rodney W. Forcade},
  title = {{Algorithm for generating derivative structures}},
  journal = {Physical Review B},
  volume = {77},
  number = {22},
  pages = {224115},
  year = {2008},
  doi = {10.1103/PhysRevB.77.224115}
}

@article{PrototypeLibrary36,
  author = {Gus L. W. Hart and Rodney W. Forcade},
  title = {{Generating derivative structures from multilattices: Algorithm and application to hcp alloys}},
  journal = {Physical Review B},
  volume = {80},
  number = {1},
  pages = {014120},
  year = {2009},
  doi = {10.1103/PhysRevB.80.014120}
}

@article{PrototypeLibrary37,
  author = {Gus L. W. Hart and Lance J. Nelson and Rodney W. Forcade},
  title = {{Generating derivative structures at a fixed concentration}},
  journal = {Computational Materials Science},
  volume = {59},
  pages = {101--107},
  year = {2012},
  doi = {10.1016/j.commatsci.2012.02.015}
}

@article{PrototypeLibrary38,
  author = {Ankit Jain and Thomas Bligaard},
  title = {{Atomic-position independent descriptor for machine learning of material properties}},
  journal = {Physical Review B},
  volume = {98},
  number = {21},
  pages = {214112},
  year = {2018},
  doi = {10.1103/PhysRevB.98.214112}
}

@article{PrototypeLibrary39,
  author = {Pavel V. Bushlanov and Vladislav A. Blatov and Artem R. Oganov},
  title = {{Topology-based crystal structure generator}},
  journal = {Computer Physics Communications},
  volume = {236},
  pages = {1--7},
  year = {2019},
  doi = {10.1016/j.cpc.2018.09.016}
}

@article{PrototypeLibrary40,
  author = {Rhys E. A. Goodall and Abhijith S. Parackal and Felix A. Faber and Rickard Armiento and Alpha A. Lee},
  title = {{Rapid discovery of stable materials by coordinate-free coarse graining}},
  journal = {Science Advances},
  volume = {8},
  number = {30},
  pages = {eabn4117},
  year = {2022},
  doi = {10.1126/sciadv.abn4117}
}

@misc{PrototypeLibrary41,
  author = {Abhijith S. Parackal and Florian Trybel and Felix Andreas Faber and Rickard Armiento},
  title = {{Screening 39 billion protostructures for materials discovery}},
  year = {2026},
  doi = {10.48550/arXiv.2601.21393},
  howpublished = {arXiv preprint arXiv:2601.21393}
}

@inproceedings{PrototypeLibrary42,
  author = {Nikita Kazeev and Wei Nong and Ignat Romanov and Ruiming Zhu and Andrey E. Ustyuzhanin and Shuya Yamazaki and Kedar Hippalgaonkar},
  title = {{Wyckoff Transformer: Generation of Symmetric Crystals}},
  booktitle = {Proceedings of the 42nd International Conference on Machine Learning},
  series = {Proceedings of Machine Learning Research},
  volume = {267},
  pages = {29495--29526},
  publisher = {PMLR},
  year = {2025},
  url = {https://proceedings.mlr.press/v267/kazeev25a.html}
}

@article{PrototypeLibrary43,
  author = {Ruiming Zhu and Wei Nong and Shuya Yamazaki and Kedar Hippalgaonkar},
  title = {{WyCryst: Wyckoff Inorganic Crystal Generator Framework}},
  journal = {Matter},
  volume = {7},
  number = {10},
  pages = {3469--3488},
  year = {2024},
  doi = {10.1016/j.matt.2024.05.042}
}

@article{PrototypeLibrary44,
  author = {Michael O’Keeffe and Maxim A. Peskov and Stuart J. Ramsden and Omar M. Yaghi},
  title = {{The Reticular Chemistry Structure Resource (RCSR) Database of, and Symbols for, Crystal Nets}},
  journal = {Accounts of Chemical Research},
  volume = {41},
  number = {12},
  pages = {1782--1789},
  year = {2008},
  doi = {10.1021/ar800124u}
}

@article{PrototypeLibrary45,
  author = {Stuart J. Ramsden and Vanessa Robins and Stephen T. Hyde},
  title = {{Three-dimensional Euclidean nets from two-dimensional hyperbolic tilings: kaleidoscopic examples}},
  journal = {Acta Crystallographica Section A: Foundations of Crystallography},
  volume = {65},
  number = {2},
  pages = {81--108},
  year = {2009},
  doi = {10.1107/S0108767308040592}
}

@article{PrototypeLibrary46,
  author = {Roald Hoffmann and Artyom A. Kabanov and Andrey A. Golov and Davide M. Proserpio},
  title = {{Homo Citans and Carbon Allotropes: For an Ethics of Citation}},
  journal = {Angewandte Chemie International Edition},
  volume = {55},
  number = {37},
  pages = {10962--10976},
  year = {2016},
  doi = {10.1002/anie.201600655}
}

@article{PrototypeLibrary47,
  author = {Michael W. Deem and Ramdas Pophale and Phillip A. Cheeseman and David J. Earl},
  title = {{Computational Discovery of New Zeolite-Like Materials}},
  journal = {The Journal of Physical Chemistry C},
  volume = {113},
  number = {51},
  pages = {21353--21360},
  year = {2009},
  doi = {10.1021/jp906984z}
}

@article{PrototypeLibrary48,
  author = {Ramdas Pophale and Phillip A. Cheeseman and Michael W. Deem},
  title = {{A database of new zeolite-like materials}},
  journal = {Physical Chemistry Chemical Physics},
  volume = {13},
  number = {27},
  pages = {12407},
  year = {2011},
  doi = {10.1039/C0CP02255A}
}

@article{PrototypeLibrary49,
  author = {Sten Haastrup and Mikkel Strange and Mohnish Pandey and Thorsten Deilmann and Per S. Schmidt and Nicki F. Hinsche and Morten N. Gjerding and Daniele Torelli and Peter M. Larsen and Anders C. Riis-Jensen and Jakob Gath and Karsten W. Jacobsen and Jens Jørgen Mortensen and Thomas Olsen and Kristian S. Thygesen},
  title = {{The Computational 2D Materials Database: high-throughput modeling and discovery of atomically thin crystals}},
  journal = {2D Materials},
  volume = {5},
  number = {4},
  pages = {042002},
  year = {2018},
  doi = {10.1088/2053-1583/aacfc1}
}

@article{PrototypeLibrary50,
  author = {M. N. Gjerding and A. Taghizadeh and A. Rasmussen and S. Ali and F. Bertoldo and T. Deilmann and U. P. Holguin and N. R. Knøsgaard and M. Kruse and S. Manti and T. G. Pedersen and T. Skovhus and M. K. Svendsen and J. J. Mortensen and T. Olsen and K. S. Thygesen},
  title = {{Recent progress of the Computational 2D Materials Database (C2DB)}},
  journal = {2D Materials},
  volume = {8},
  number = {4},
  pages = {044002},
  year = {2021},
  doi = {10.1088/2053-1583/ac1059}
}

@article{PrototypeLibrary51,
  author = {Giancarlo Trimarchi and Xiuwen Zhang and Michael J. DeVries Vermeer and Jacqueline Cantwell and Kenneth R. Poeppelmeier and Alex Zunger},
  title = {{Emergence of a few distinct structures from a single formal structure type during high-throughput screening for stable compounds: The case of RbCuS and RbCuSe}},
  journal = {Physical Review B},
  volume = {92},
  number = {16},
  pages = {165103},
  year = {2015},
  doi = {10.1103/PhysRevB.92.165103}
}

@article{PrototypeUse01,
  author = {Nathan J. Szymanski and Christopher J. Bartel},
  title = {{Establishing baselines for generative discovery of inorganic crystals}},
  journal = {Materials Horizons},
  volume = {12},
  number = {19},
  pages = {8000--8011},
  year = {2025},
  doi = {10.1039/d5mh00010f}
}

@article{PrototypeUse02,
  author = {Lai Wei and Nihang Fu and Edirisuriya M. Dilanga Siriwardane and Wenhui Yang and Sadman Sadeed Omee and Rongzhi Dong and Rui Xin and Jianjun Hu},
  title = {{TCSP: a Template-Based Crystal Structure Prediction Algorithm for Materials Discovery}},
  journal = {Inorganic Chemistry},
  volume = {61},
  number = {22},
  pages = {8431--8439},
  year = {2022},
  doi = {10.1021/acs.inorgchem.1c03879}
}

@article{PrototypeUse03,
  author = {Minoru Kusaba and Chang Liu and Ryo Yoshida},
  title = {{Crystal structure prediction with machine learning-based element substitution}},
  journal = {Computational Materials Science},
  volume = {211},
  pages = {111496},
  year = {2022},
  doi = {10.1016/j.commatsci.2022.111496}
}

@article{PrototypeUse04,
  author = {Lai Wei and Rong-Zhi Dong and Ni-Hang Fu and Sadman Sadeed Omee and Jian-Jun Hu},
  title = {{TCSP 2.0: Template based crystal structure prediction with improved oxidation state prediction and chemistry heuristics}},
  journal = {Computational Materials Science},
  volume = {261},
  pages = {114317},
  year = {2026},
  doi = {10.1016/j.commatsci.2025.114317}
}

@article{PrototypeUse05,
  author = {Cem Oran and Riccarda Caputo and Pierre Villars and Adem Tekin},
  title = {{PNcsp+: A Periodic Number-Based Crystal Structure Prediction Method Enhanced by Machine Learning}},
  journal = {Journal of Chemical Theory and Computation},
  volume = {22},
  number = {7},
  pages = {3761--3771},
  year = {2026},
  doi = {10.1021/acs.jctc.6c00044}
}

@article{PrototypeUse06,
  author = {Sean D. Griesemer and Logan T. Ward and Chris M. Wolverton},
  title = {{High-throughput crystal structure solution using prototypes}},
  journal = {Physical Review Materials},
  volume = {5},
  number = {10},
  pages = {105003},
  year = {2021},
  doi = {10.1103/physrevmaterials.5.105003}
}

@article{PrototypeUse07,
  author = {Abhijith S. Parackal and Rhys E. A. Goodall and Felix A. Faber and Rickard Armiento},
  title = {{Identifying crystal structures beyond known prototypes from x-ray powder diffraction spectra}},
  journal = {Physical Review Materials},
  volume = {8},
  number = {10},
  pages = {103801},
  year = {2024},
  doi = {10.1103/physrevmaterials.8.103801}
}

@article{PrototypeUse08,
  author = {Amil Merchant and Simon Batzner and S. Schoenholz and M. Aykol and Gowoon Cheon and E. Cubuk},
  title = {{Scaling deep learning for materials discovery}},
  journal = {Nature},
  volume = {624},
  number = {7990},
  pages = {80--85},
  year = {2023},
  doi = {10.1038/s41586-023-06735-9}
}

@article{PrototypeUse09,
  author = {Tri Minh Nguyen and Sherif Abdulkader Tawfik and Truyen Tran and Sunil Gupta and Santu Rana and Svetha Venkatesh},
  title = {{Efficient symmetry-aware materials generation via hierarchical generative flow networks}},
  journal = {Digital Discovery},
  volume = {5},
  number = {2},
  pages = {686--697},
  year = {2026},
  doi = {10.1039/d4dd00392f}
}

@article{PrototypeUse10,
  author = {Takanori Ishii and Kaoru Hisama and Kohei Shinohara},
  title = {{Symmetry-aware conditional generation of crystal structures using diffusion models}},
  journal = {Computational Materials Science},
  volume = {271},
  pages = {114789},
  year = {2026},
  doi = {10.1016/j.commatsci.2026.114789}
}

@misc{PrototypeUse11,
  author = {Masahiro Negishi and Aron Walsh},
  title = {{Substitution-Based Analysis of Structural Novelty for Generative Models of Materials}},
  year = {2026},
  doi = {10.48550/arxiv.2606.23166},
  howpublished = {arXiv preprint arXiv:2606.23166}
}

@article{PrototypeUse12,
  author = {Rongzhi Dong and Yuqi Song and Edirisuriya M. Dilanga Siriwardane and Jianjun Hu},
  title = {{Discovery of 2D Materials using Transformer Network‐Based Generative Design}},
  journal = {Advanced Intelligent Systems},
  volume = {5},
  number = {12},
  pages = {2300141},
  year = {2023},
  doi = {10.1002/aisy.202300141}
}

@article{PrototypeUse13,
  author = {Peder Lyngby and Kristian Sommer Thygesen},
  title = {{Data-driven discovery of 2D materials by deep generative models}},
  journal = {npj Computational Materials},
  volume = {8},
  number = {1},
  pages = {232},
  year = {2022},
  doi = {10.1038/s41524-022-00923-3}
}

@article{PrototypeUse14,
  author = {Zhen-Xiong Shen and Chuanxun Su and Lixin He},
  title = {{High-throughput computation and structure prototype analysis for two-dimensional ferromagnetic materials}},
  journal = {npj Computational Materials},
  volume = {8},
  number = {1},
  pages = {132},
  year = {2022},
  doi = {10.1038/s41524-022-00813-8}
}

@article{PrototypeUse15,
  author = {Raul A. Flores and Christopher Paolucci and Kirsten T. Winther and Ankit Jain and José Antonio Garrido Torres and Muratahan Aykol and Joseph H. Montoya and Jens K. Nørskov and Michal Bajdich and Thomas Bligaard},
  title = {{Active Learning Accelerated Discovery of Stable Iridium Oxide Polymorphs for the Oxygen Evolution Reaction}},
  journal = {Chemistry of Materials},
  volume = {32},
  number = {13},
  pages = {5854--5863},
  year = {2020},
  doi = {10.1021/acs.chemmater.0c01894}
}

@article{PrototypeUse16,
  author = {Andrij Vasylenko and Dmytro Antypov and Sven Schewe and Luke M. Daniels and John B. Claridge and Matthew Stephen Dyer and Matthew J. Rosseinsky},
  title = {{Digital features of chemical elements extracted from local geometries in crystal structures}},
  journal = {Digital Discovery},
  volume = {4},
  number = {2},
  pages = {477--485},
  year = {2025},
  doi = {10.1039/d4dd00346b}
}

@article{PrototypeUse17,
  author = {Yawen Li and Jingxiu Yang and Ruoting Zhao and Yilin Zhang and Xinjiang Wang and Xin Hai He and Yuhao Fu and Lijun Zhang},
  title = {{Design of Organic--Inorganic Hybrid Heterostructured Semiconductors via High-Throughput Materials Screening for Optoelectronic Applications}},
  journal = {Journal of the American Chemical Society},
  volume = {144},
  number = {36},
  pages = {16656--16666},
  year = {2022},
  doi = {10.1021/jacs.2c07434}
}

@article{PrototypeUse18,
  author = {Pranab Sarker and Tyler Harrington and Cormac Toher and Corey Oses and Mojtaba Samiee and Jon‐Paul Maria and Donald W. Brenner and Kenneth S. Vecchio and Stefano Curtarolo},
  title = {{High-entropy high-hardness metal carbides discovered by entropy descriptors}},
  journal = {Nature Communications},
  volume = {9},
  number = {1},
  pages = {4980},
  year = {2018},
  doi = {10.1038/s41467-018-07160-7}
}

@article{PrototypeUse19,
  author = {Simon Divilov and Hagen Eckert and David Hicks and Corey Oses and Cormac Toher and Rico Friedrich and Marco Esters and Michael J. Mehl and Adam C. Zettel and Yoav Lederer and Eva Zurek and Jon‐Paul Maria and Donald W. Brenner and Xiomara Campilongo and Suzana Filipović and William G. Fahrenholtz and Caillin J. Ryan and Christopher M. DeSalle and Ryan J. Crealese and Douglas E. Wolfe and Arrigo Calzolari and Stefano Curtarolo},
  title = {{Disordered enthalpy--entropy descriptor for high-entropy ceramics discovery}},
  journal = {Nature},
  volume = {625},
  number = {7993},
  pages = {66--73},
  year = {2024},
  doi = {10.1038/s41586-023-06786-y}
}

@article{PrototypeUse20,
  author = {Yamil J. Colón and Diego A. Gómez‐Gualdrón and Randall Q. Snurr},
  title = {{Topologically Guided, Automated Construction of Metal--Organic Frameworks and Their Evaluation for Energy-Related Applications}},
  journal = {Crystal Growth \& Design},
  volume = {17},
  number = {11},
  pages = {5801--5810},
  year = {2017},
  doi = {10.1021/acs.cgd.7b00848}
}

@article{PrototypeUse21,
  author = {Hyunsoo Park and Sauradeep Majumdar and Xiaoqi Zhang and Jihan Kim and Berend Smit},
  title = {{Inverse design of metal--organic frameworks for direct air capture of CO$_2$ via deep reinforcement learning}},
  journal = {Digital Discovery},
  volume = {3},
  number = {4},
  pages = {728--741},
  year = {2024},
  doi = {10.1039/d4dd00010b}
}

@article{PrototypeUse22,
  author = {Geoffroy Hautier and Chris Fischer and Virginie Ehrlacher and Anubhav Jain and Gerbrand Ceder},
  title = {{Data Mined Ionic Substitutions for the Discovery of New Compounds}},
  journal = {Inorganic Chemistry},
  volume = {50},
  number = {2},
  pages = {656--663},
  year = {2011},
  doi = {10.1021/ic102031h}
}

@article{PrototypeUse23,
  author = {Geoffroy Hautier and Christopher C. Fischer and Anubhav Jain and Tim Mueller and Gerbrand Ceder},
  title = {{Finding Nature’s Missing Ternary Oxide Compounds Using Machine Learning and Density Functional Theory}},
  journal = {Chemistry of Materials},
  volume = {22},
  number = {12},
  pages = {3762--3767},
  year = {2010},
  doi = {10.1021/cm100795d}
}

@article{ProteinFold01,
  author = {Cyrus Chothia},
  title = {{One thousand families for the molecular biologist}},
  journal = {Nature},
  volume = {357},
  number = {6379},
  pages = {543--544},
  year = {1992},
  doi = {10.1038/357543a0}
}

@article{ProteinFold02,
  author = {Chao Zhang and Charles DeLisi},
  title = {{Estimating the number of protein folds}},
  journal = {Journal of Molecular Biology},
  volume = {284},
  number = {5},
  pages = {1301--1305},
  year = {1998},
  doi = {10.1006/jmbi.1998.2282}
}

@article{ProteinFold03,
  author = {Sridhar Govindarajan and Ruben Recabarren and Richard A. Goldstein},
  title = {{Estimating the total number of protein folds}},
  journal = {Proteins: Structure, Function, and Genetics},
  volume = {35},
  number = {4},
  pages = {408--414},
  year = {1999},
  doi = {10.1002/(sici)1097-0134(19990601)35:4<408::aid-prot4>3.0.co;2-a}
}

@article{ProteinFold04,
  author = {Yang Zhang and Jeffrey Skolnick},
  title = {{The protein structure prediction problem could be solved using the current PDB library}},
  journal = {Proceedings of the National Academy of Sciences},
  volume = {102},
  number = {4},
  pages = {1029--1034},
  year = {2005},
  doi = {10.1073/pnas.0407152101}
}

@article{ProteinFold05,
  author = {Yang Zhang and Isaac A. Hubner and Adrián K. Arakaki and Eugene I. Shakhnovich and Jeffrey Skolnick},
  title = {{On the origin and highly likely completeness of single-domain protein structures}},
  journal = {Proceedings of the National Academy of Sciences},
  volume = {103},
  number = {8},
  pages = {2605--2610},
  year = {2006},
  doi = {10.1073/pnas.0509379103}
}

@article{ProteinFold07,
  author = {Sergey Nepomnyachiy and Nir Ben-Tal and Rachel Kolodny},
  title = {{Global view of the protein universe}},
  journal = {Proceedings of the National Academy of Sciences},
  volume = {111},
  number = {32},
  pages = {11691--11696},
  year = {2014},
  doi = {10.1073/pnas.1403395111}
}

@article{ProteinAI01,
  author = {Nicola Bordin and Ian Sillitoe and Vamsi Nallapareddy and Clemens Rauer and Su Datt Lam and Vaishali P. Waman and Neeladri Sen and Michael Heinzinger and Maria Littmann and Stephanie Kim and Sameer Velankar and Martin Steinegger and Burkhard Rost and Christine Orengo},
  title = {{AlphaFold2 reveals commonalities and novelties in protein structure space for 21 model organisms}},
  journal = {Communications Biology},
  volume = {6},
  number = {1},
  pages = {160},
  year = {2023},
  doi = {10.1038/s42003-023-04488-9}
}

@article{ProteinAI04,
  author = {Joseph L. Watson and David Juergens and Nathaniel R. Bennett and Brian L. Trippe and Jason Yim and Helen E. Eisenach and Woody Ahern and Andrew J. Borst and Robert J. Ragotte and Lukas F. Milles and Basile I. M. Wicky and Nikita Hanikel and Samuel J. Pellock and Alexis Courbet and William Sheffler and Jue Wang and Preetham Venkatesh and Isaac Sappington and Susana Vázquez Torres and Anna Lauko and Valentin De Bortoli and Emile Mathieu and Sergey Ovchinnikov and Regina Barzilay and Tommi S. Jaakkola and Frank DiMaio and Minkyung Baek and David Baker},
  title = {{De novo design of protein structure and function with RFdiffusion}},
  journal = {Nature},
  volume = {620},
  number = {7976},
  pages = {1089--1100},
  year = {2023},
  doi = {10.1038/s41586-023-06415-8}
}

@article{ProteinAI05,
  author = {J. Dauparas and I. Anishchenko and N. Bennett and H. Bai and R. J. Ragotte and L. F. Milles and B. I. M. Wicky and A. Courbet and R. J. de Haas and N. Bethel and P. J. Y. Leung and T. F. Huddy and S. Pellock and D. Tischer and F. Chan and B. Koepnick and H. Nguyen and A. Kang and B. Sankaran and A. K. Bera and N. P. King and D. Baker},
  title = {{Robust deep learning--based protein sequence design using ProteinMPNN}},
  journal = {Science},
  volume = {378},
  number = {6615},
  pages = {49--56},
  year = {2022},
  doi = {10.1126/science.add2187}
}

@article{ProteinAI06,
  author = {Inigo Barrio‐Hernandez and Jingi Yeo and Jürgen Jänes and Milot Mirdita and Cameron L. M. Gilchrist and Tanita Wein and Mihály Váradi and Sameer S Velankar and Pedro Beltrão and Martin Steinegger},
  title = {{Clustering predicted structures at the scale of the known protein universe}},
  journal = {Nature},
  volume = {622},
  number = {7983},
  pages = {637--645},
  year = {2023},
  doi = {10.1038/s41586-023-06510-w}
}

@article{ProteinAI07,
  author = {Janani Durairaj and Andrew M. Waterhouse and Toomas Mets and Tetiana Brodiazhenko and Minhal Abdullah and Gabriel Studer and Gerardo Tauriello and Mehmet Akdel and Antonina Andreeva and Alex Bateman and Tanel Tenson and Vasili Hauryliuk and Torsten Schwede and Joana Pereira},
  title = {{Uncovering new families and folds in the natural protein universe}},
  journal = {Nature},
  volume = {622},
  number = {7983},
  pages = {646--653},
  year = {2023},
  doi = {10.1038/s41586-023-06622-3}
}

@article{PrototypeCase01,
  author = {Marina R. Filip and Feliciano Giustino},
  title = {{The geometric blueprint of perovskites}},
  journal = {Proceedings of the National Academy of Sciences},
  volume = {115},
  number = {21},
  pages = {5397--5402},
  year = {2018},
  doi = {10.1073/pnas.1719179115}
}

@article{PrototypeCase02,
  author = {Lanhua Wei and P. K. Kuo and R. L. Thomas and T. R. Anthony and W. F. Banholzer},
  title = {{Thermal conductivity of isotopically modified single crystal diamond}},
  journal = {Physical Review Letters},
  volume = {70},
  number = {24},
  pages = {3764--3767},
  year = {1993},
  doi = {10.1103/physrevlett.70.3764}
}

@article{PrototypeCase03,
  author = {Joon Sang Kang and Man Li and Huan Wu and Huuduy Nguyen and Yongjie Hu},
  title = {{Experimental observation of high thermal conductivity in boron arsenide}},
  journal = {Science},
  volume = {361},
  number = {6402},
  pages = {575--578},
  year = {2018},
  doi = {10.1126/science.aat5522}
}

@article{PrototypeCase04,
  author = {Ramaswamy Murugan and Venkataraman Thangadurai and Werner Weppner},
  title = {{Fast Lithium Ion Conduction in Garnet-Type Li$_7$La$_3$Zr$_2$O$_{12}$}},
  journal = {Angewandte Chemie International Edition},
  volume = {46},
  number = {41},
  pages = {7778--7781},
  year = {2007},
  doi = {10.1002/anie.200701144}
}

@article{PrototypeCase05,
  author = {A. K. Padhi and K. S. Nanjundaswamy and J. B. Goodenough},
  title = {{Phospho-olivines as Positive-Electrode Materials for Rechargeable Lithium Batteries}},
  journal = {Journal of The Electrochemical Society},
  volume = {144},
  number = {4},
  pages = {1188--1194},
  year = {1997},
  doi = {10.1149/1.1837571}
}

@article{PrototypeCase06,
  author = {J.B. Goodenough and H.Y-P. Hong and J.A. Kafalas},
  title = {{Fast Na+-ion transport in skeleton structures}},
  journal = {Materials Research Bulletin},
  volume = {11},
  number = {2},
  pages = {203--220},
  year = {1976},
  doi = {10.1016/0025-5408(76)90077-5}
}

@article{PrototypeCase07,
  author = {E. S. Grew and A. J. Locock and S. J. Mills and I. O. Galuskina and E. V. Galuskin and U. Halenius},
  title = {{Nomenclature of the garnet supergroup}},
  journal = {American Mineralogist},
  volume = {98},
  number = {4},
  pages = {785--811},
  year = {2013},
  doi = {10.2138/am.2013.4201}
}

@article{PrototypeCase08,
  author = {Roger H. Mitchell and Mark D. Welch and Anton R. Chakhmouradian},
  title = {{Nomenclature of the perovskite supergroup: A hierarchical system of classification based on crystal structure and composition}},
  journal = {Mineralogical Magazine},
  volume = {81},
  number = {3},
  pages = {411--461},
  year = {2017},
  doi = {10.1180/minmag.2016.080.156}
}

@article{PrototypeCase09,
  author = {Ferdinando Bosi and Cristian Biagioni and Marco Pasero},
  title = {{Nomenclature and classification of the spinel supergroup}},
  journal = {European Journal of Mineralogy},
  volume = {31},
  number = {1},
  pages = {183--192},
  year = {2019},
  doi = {10.1127/ejm/2019/0031-2788}
}

@article{PrototypeCase10,
  author = {Julie A. Bowles and Mike J. Jackson and Thelma S. Berquo and Peter A. Slheid and Jeffrey S. Gee},
  title = {{Inferred time- and temperature-dependent cation ordering in natural titanomagnetites}},
  journal = {Nature Communications},
  volume = {4},
  number = {1},
  pages = {1916},
  year = {2013},
  doi = {10.1038/ncomms2938}
}

@incollection{MineralFamily01,
  author = {Ribbe, P. H.},
  editor = {Ribbe, P. H.},
  title = {{The chemistry, structure and nomenclature of feldspars}},
  booktitle = {Feldspar Mineralogy},
  series = {Reviews in Mineralogy},
  volume = {2},
  pages = {1--20},
  edition = {Second},
  publisher = {De Gruyter},
  address = {Berlin, Boston},
  year = {1983},
  doi = {10.1515/9781501508547-006}
}

@misc{MineralFamily02,
  editor = {John W. Anthony and Richard A. Bideaux and Kenneth W. Bladh and Monte C. Nichols},
  title = {{Handbook of Mineralogy}},
  url = {https://handbookofmineralogy.org/pdf-search/},
  howpublished = {Mineralogical Society of America, Chantilly, VA},
  note = {Online edition, accessed 11 October 2026}
}

@article{CSPReview01,
  author = {Oganov, Artem R. and Pickard, Chris J. and Zhu, Qiang and Needs, Richard J.},
  title = {{Structure prediction drives materials discovery}},
  journal = {Nature Reviews Materials},
  volume = {4},
  number = {5},
  pages = {331--348},
  year = {2019},
  doi = {10.1038/s41578-019-0101-8}
}

@article{CSPReview02,
  author = {De Breuck, Pierre-Paul and Wang, Hai-Chen and Rignanese, Gian-Marco and Botti, Silvana and Marques, Miguel A. L.},
  title = {{Generative AI for crystal structures: a review}},
  journal = {npj Computational Materials},
  volume = {11},
  number = {1},
  pages = {370},
  year = {2025},
  doi = {10.1038/s41524-025-01881-2}
}

@article{DatasetMP,
  author = {Jain, Anubhav and Ong, Shyue Ping and Hautier, Geoffroy and Chen, Wei and Richards, William Davidson and Dacek, Stephen and Cholia, Shreyas and Gunter, Dan and Skinner, David and Ceder, Gerbrand and Persson, Kristin A.},
  title = {{Commentary: The Materials Project: A materials genome approach to accelerating materials innovation}},
  journal = {APL Materials},
  volume = {1},
  number = {1},
  pages = {011002},
  year = {2013},
  doi = {10.1063/1.4812323}
}

@article{DatasetOQMD,
  author = {Saal, James E. and Kirklin, Scott and Aykol, Muratahan and Meredig, Bryce and Wolverton, C.},
  title = {{Materials Design and Discovery with High-Throughput Density Functional Theory: The Open Quantum Materials Database (OQMD)}},
  journal = {JOM},
  volume = {65},
  number = {11},
  pages = {1501--1509},
  year = {2013},
  doi = {10.1007/s11837-013-0755-4}
}

@misc{DatasetLeMat,
  author = {{LeMaterial}},
  title = {{LeMat-Bulk Dataset}},
  year = {2024},
  doi = {10.57967/hf/3762},
  howpublished = {Hugging Face dataset}
}

@article{Alexandrov2011UnderlyingNets,
  author = {E. V. Alexandrov and V. A. Blatov and A. V. Kochetkov and D. M. Proserpio},
  title = {{Underlying nets in three-periodic coordination polymers: topology, taxonomy and prediction from a computer-aided analysis of the Cambridge Structural Database}},
  journal = {CrystEngComm},
  volume = {13},
  number = {12},
  pages = {3947},
  year = {2011},
  doi = {10.1039/c0ce00636j}
}

@article{Antunes2024CrystaLLM,
  author = {Luis M. Antunes and Keith T. Butler and Ricardo Grau-Crespo},
  title = {{Crystal structure generation with autoregressive large language modeling}},
  journal = {Nature Communications},
  volume = {15},
  number = {1},
  pages = {10570},
  year = {2024},
  doi = {10.1038/s41467-024-54639-7}
}

@article{Aykol2019Network,
  author = {Muratahan Aykol and Vinay I. Hegde and Linda Hung and Santosh Suram and Patrick Herring and Chris Wolverton and Jens S. Hummelshøj},
  title = {{Network analysis of synthesizable materials discovery}},
  journal = {Nature Communications},
  volume = {10},
  number = {1},
  pages = {2018},
  year = {2019},
  doi = {10.1038/s41467-019-10030-5}
}

@article{Baird2024MPTS,
  author = {Sterling G. Baird and Hasan M. Sayeed and Joseph Montoya and Taylor D. Sparks},
  title = {{matbench-genmetrics: A Python library for benchmarking crystal structure generative models using time-based splits of Materials Project structures}},
  journal = {Journal of Open Source Software},
  volume = {9},
  number = {97},
  pages = {5618},
  year = {2024},
  doi = {10.21105/joss.05618}
}

@article{Bartok2013SOAP,
  author = {Albert P. Bartók and Risi Kondor and Gábor Csányi},
  title = {{On representing chemical environments}},
  journal = {Physical Review B},
  volume = {87},
  number = {18},
  pages = {184115},
  year = {2013},
  doi = {10.1103/physrevb.87.184115}
}

@article{Blatov2014ToposPro,
  author = {Vladislav A. Blatov and Alexander P. Shevchenko and Davide M. Proserpio},
  title = {{Applied Topological Analysis of Crystal Structures with the Program Package ToposPro}},
  journal = {Crystal Growth \& Design},
  volume = {14},
  number = {7},
  pages = {3576--3586},
  year = {2014},
  doi = {10.1021/cg500498k}
}

@article{Cheetham2024GNoME,
  author = {Anthony K. Cheetham and Ram Seshadri},
  title = {{Artificial Intelligence Driving Materials Discovery? Perspective on the Article: Scaling Deep Learning for Materials Discovery}},
  journal = {Chemistry of Materials},
  volume = {36},
  number = {8},
  pages = {3490--3495},
  year = {2024},
  doi = {10.1021/acs.chemmater.4c00643}
}

@article{Curtarolo2012AFLOW,
  author = {Stefano Curtarolo and Wahyu Setyawan and Gus L.W. Hart and Michal Jahnatek and Roman V. Chepulskii and Richard H. Taylor and Shidong Wang and Junkai Xue and Kesong Yang and Ohad Levy and Michael J. Mehl and Harold T. Stokes and Denis O. Demchenko and Dane Morgan},
  title = {{AFLOW: An automatic framework for high-throughput materials discovery}},
  journal = {Computational Materials Science},
  volume = {58},
  pages = {218--226},
  year = {2012},
  doi = {10.1016/j.commatsci.2012.02.005}
}

@article{Curtarolo2012AFLOWLIB,
  author = {Stefano Curtarolo and Wahyu Setyawan and Shidong Wang and Junkai Xue and Kesong Yang and Richard H. Taylor and Lance J. Nelson and Gus L.W. Hart and Stefano Sanvito and Marco Buongiorno-Nardelli and Natalio Mingo and Ohad Levy},
  title = {{AFLOWLIB.ORG: A distributed materials properties repository from high-throughput ab initio calculations}},
  journal = {Computational Materials Science},
  volume = {58},
  pages = {227--235},
  year = {2012},
  doi = {10.1016/j.commatsci.2012.02.002}
}

@article{Daams2000AET,
  author = {Jo Daams and Pierre Villars},
  title = {{Atomic environments in relation to compound prediction}},
  journal = {Engineering Applications of Artificial Intelligence},
  volume = {13},
  number = {5},
  pages = {507--511},
  year = {2000},
  doi = {10.1016/s0952-1976(00)00029-4}
}

@article{Davies2016Screening,
  author = {Daniel W. Davies and Keith T. Butler and Adam J. Jackson and Andrew Morris and Jarvist M. Frost and Jonathan M. Skelton and Aron Walsh},
  title = {{Computational Screening of All Stoichiometric Inorganic Materials}},
  journal = {Chem},
  volume = {1},
  number = {4},
  pages = {617--627},
  year = {2016},
  doi = {10.1016/j.chempr.2016.09.010}
}

@article{Dshemuchadse2015Statistics,
  author = {Julia Dshemuchadse and Walter Steurer},
  title = {{Some Statistics on Intermetallic Compounds}},
  journal = {Inorganic Chemistry},
  volume = {54},
  number = {3},
  pages = {1120--1128},
  year = {2015},
  doi = {10.1021/ic5024482}
}

@article{Grazulis2009COD,
  author = {Saulius Gražulis and Daniel Chateigner and Robert T. Downs and A. F. T. Yokochi and Miguel Quirós and Luca Lutterotti and Elena Manakova and Justas Butkus and Peter Moeck and Armel Le Bail},
  title = {{Crystallography Open Database -- an open-access collection of crystal structures}},
  journal = {Journal of Applied Crystallography},
  volume = {42},
  number = {4},
  pages = {726--729},
  year = {2009},
  doi = {10.1107/s0021889809016690}
}

@article{Grazulis2012COD,
  author = {Saulius Gražulis and Adriana Daškevič and Andrius Merkys and Daniel Chateigner and Luca Lutterotti and Miguel Quirós and Nadezhda R. Serebryanaya and Peter Moeck and Robert T. Downs and Armel Le Bail},
  title = {{Crystallography Open Database (COD): an open-access collection of crystal structures and platform for world-wide collaboration}},
  journal = {Nucleic Acids Research},
  volume = {40},
  number = {D1},
  pages = {D420--D427},
  year = {2012},
  doi = {10.1093/nar/gkr900}
}

@article{Hellenbrandt2004ICSD,
  author = {Mariette Hellenbrandt},
  title = {{The Inorganic Crystal Structure Database (ICSD)---Present and Future}},
  journal = {Crystallography Reviews},
  volume = {10},
  number = {1},
  pages = {17--22},
  year = {2004},
  doi = {10.1080/08893110410001664882}
}

@article{Hicks2018AflowSym,
  author = {David Hicks and Corey Oses and Eric Gossett and Geena Gomez and Richard H. Taylor and Cormac Toher and Michael J. Mehl and Ohad Levy and Stefano Curtarolo},
  title = {{AFLOW-SYM: platform for the complete, automatic and self-consistent symmetry analysis of crystals}},
  journal = {Acta Crystallographica Section A: Foundations and Advances},
  volume = {74},
  number = {3},
  pages = {184--203},
  year = {2018},
  doi = {10.1107/s2053273318003066}
}

@article{Kirklin2015OQMD,
  author = {Scott Kirklin and James E Saal and Bryce Meredig and Alex Thompson and Jeff W Doak and Muratahan Aykol and Stephan Rühl and Chris Wolverton},
  title = {{The Open Quantum Materials Database (OQMD): assessing the accuracy of DFT formation energies}},
  journal = {npj Computational Materials},
  volume = {1},
  number = {1},
  pages = {15010},
  year = {2015},
  doi = {10.1038/npjcompumats.2015.10}
}

@article{Leeman2024Challenges,
  author = {Josh Leeman and Yuhan Liu and Joseph Stiles and Scott B. Lee and Prajna Bhatt and Leslie M. Schoop and Robert G. Palgrave},
  title = {{Challenges in High-Throughput Inorganic Materials Prediction and Autonomous Synthesis}},
  journal = {PRX Energy},
  volume = {3},
  number = {1},
  pages = {011002},
  year = {2024},
  doi = {10.1103/prxenergy.3.011002}
}

@article{Ong2013Pymatgen,
  author = {Shyue Ping Ong and William Davidson Richards and Anubhav Jain and Geoffroy Hautier and Michael Kocher and Shreyas Cholia and Dan Gunter and Vincent L. Chevrier and Kristin A. Persson and Gerbrand Ceder},
  title = {{Python Materials Genomics (pymatgen): A robust, open-source python library for materials analysis}},
  journal = {Computational Materials Science},
  volume = {68},
  pages = {314--319},
  year = {2013},
  doi = {10.1016/j.commatsci.2012.10.028}
}

@article{Pan2021CrystalNN,
  author = {Hillary Pan and Alex M. Ganose and Matthew Horton and Muratahan Aykol and Kristin A. Persson and Nils E. R. Zimmermann and Anubhav Jain},
  title = {{Benchmarking Coordination Number Prediction Algorithms on Inorganic Crystal Structures}},
  journal = {Inorganic Chemistry},
  volume = {60},
  number = {3},
  pages = {1590--1603},
  year = {2021},
  doi = {10.1021/acs.inorgchem.0c02996}
}

@article{Pauling1929Principles,
  author = {Linus Pauling},
  title = {{The principles determining the structure of complex ionic crystals}},
  journal = {Journal of the American Chemical Society},
  volume = {51},
  number = {4},
  pages = {1010--1026},
  year = {1929},
  doi = {10.1021/ja01379a006}
}

@article{Riebesell2025Matbench,
  author = {Janosh Riebesell and Rhys E. A. Goodall and Philipp Benner and Yuan Chiang and Bowen Deng and Gerbrand Ceder and Mark Asta and Alpha A. Lee and Anubhav Jain and Kristin A. Persson},
  title = {{A framework to evaluate machine learning crystal stability predictions}},
  journal = {Nature Machine Intelligence},
  volume = {7},
  number = {6},
  pages = {836--847},
  year = {2025},
  doi = {10.1038/s42256-025-01055-1}
}

@article{Schmidt2023Alexandria,
  author = {Jonathan Schmidt and Noah Hoffmann and Hai‐Chen Wang and Pedro Borlido and Pedro J. M. A. Carriço and Tiago F. T. Cerqueira and Silvana Botti and Miguel A. L. Marques},
  title = {{Machine-Learning-Assisted Determination of the Global Zero-Temperature Phase Diagram of Materials}},
  journal = {Advanced Materials},
  volume = {35},
  number = {22},
  pages = {2210788},
  year = {2023},
  doi = {10.1002/adma.202210788}
}

@article{Schmidt2024Alexandria,
  author = {Jonathan Schmidt and Tiago F.T. Cerqueira and Aldo H. Romero and Antoine Loew and Fabian Jäger and Hai-Chen Wang and Silvana Botti and Miguel A.L. Marques},
  title = {{Improving machine-learning models in materials science through large datasets}},
  journal = {Materials Today Physics},
  volume = {48},
  pages = {101560},
  year = {2024},
  doi = {10.1016/j.mtphys.2024.101560}
}

@article{Sickafus1999Spinel,
  author = {Kurt E. Sickafus and John M. Wills and Norman W. Grimes},
  title = {{Structure of Spinel}},
  journal = {Journal of the American Ceramic Society},
  volume = {82},
  number = {12},
  pages = {3279--3292},
  year = {1999},
  doi = {10.1111/j.1151-2916.1999.tb02241.x}
}

@article{Talirz2020MaterialsCloud,
  author = {Leopold Talirz and Snehal Kumbhar and Elsa Passaro and Aliaksandr V. Yakutovich and Valeria Granata and Fernando Gargiulo and Marco Borelli and Martin Uhrin and Sebastiaan P. Huber and Spyros Zoupanos and Carl S. Adorf and Casper Welzel Andersen and Ole Schütt and Carlo A. Pignedoli and Daniele Passerone and Joost VandeVondele and Thomas C. Schulthess and Berend Smit and Giovanni Pizzi and Nicola Marzari},
  title = {{Materials Cloud, a platform for open computational science}},
  journal = {Scientific Data},
  volume = {7},
  number = {1},
  pages = {299},
  year = {2020},
  doi = {10.1038/s41597-020-00637-5}
}

@article{Togo2024Spglib,
  author = {Atsushi Togo and Kohei Shinohara and Isao Tanaka},
  title = {{Spglib: a software library for crystal symmetry search}},
  journal = {Science and Technology of Advanced Materials: Methods},
  volume = {4},
  number = {1},
  pages = {2384822},
  year = {2024},
  doi = {10.1080/27660400.2024.2384822}
}

@article{Urusov2009SpaceGroups,
  author = {V. S. Urusov and T. N. Nadezhina},
  title = {{Frequency distribution and selection of space groups in inorganic crystal chemistry}},
  journal = {Journal of Structural Chemistry},
  volume = {50},
  number = {S1},
  pages = {22--37},
  year = {2009},
  doi = {10.1007/s10947-009-0186-9}
}

@article{Villars2004AET,
  author = {P. Villars and K. Cenzual and J. Daams and Y. Chen and S. Iwata},
  title = {{Data-driven atomic environment prediction for binaries using the Mendeleev number}},
  journal = {Journal of Alloys and Compounds},
  volume = {367},
  number = {1-2},
  pages = {167--175},
  year = {2004},
  doi = {10.1016/j.jallcom.2003.08.060}
}

@incollection{Villars2019PaulingFile,
  author = {Pierre Villars and Karin Cenzual and Roman Gladyshevskii and Shuichi Iwata},
  title = {{Pauling File: Toward a Holistic View}},
  booktitle = {Materials Informatics: Methods, Tools and Applications},
  pages = {55--106},
  publisher = {Wiley-VCH},
  address = {Weinheim},
  year = {2019},
  doi = {10.1002/9783527802265.ch3}
}

@article{Widdowson2022AMD,
  author = {Daniel Widdowson and Marco M. Mosca and Angeles Pulido and Andrew I. Cooper and Vitaliy Kurlin},
  title = {{Average minimum distances of periodic point sets -- foundational invariants for mapping periodic crystals}},
  journal = {MATCH Communications in Mathematical and in Computer Chemistry},
  volume = {87},
  number = {3},
  pages = {529--559},
  year = {2022},
  doi = {10.46793/match.87-3.529w}
}

@article{Zeni2025MatterGen,
  author = {Claudio Zeni and Robert Pinsler and Daniel Zügner and Andrew Fowler and Matthew Horton and Xiang Fu and Zilong Wang and Aliaksandra Shysheya and Jonathan Crabbé and Shoko Ueda and Roberto Sordillo and Lixin Sun and Jake Smith and Bichlien Nguyen and Hannes Schulz and Sarah Lewis and Chin-Wei Huang and Ziheng Lu and Yichi Zhou and Han Yang and Hongxia Hao and Jielan Li and Chunlei Yang and Wenjie Li and Ryota Tomioka and Tian Xie},
  title = {{A generative model for inorganic materials design}},
  journal = {Nature},
  volume = {639},
  number = {8055},
  pages = {624--632},
  year = {2025},
  doi = {10.1038/s41586-025-08628-5}
}

@inproceedings{Xie2022CDVAE,
  author = {Tian Xie and Xiang Fu and Octavian-Eugen Ganea and Regina Barzilay and Tommi Jaakkola},
  title = {{Crystal Diffusion Variational Autoencoder for Periodic Material Generation}},
  booktitle = {International Conference on Learning Representations (ICLR)},
  year = {2022},
  url = {https://arxiv.org/abs/2110.06197}
}

@inproceedings{Jiao2023DiffCSP,
  author = {Rui Jiao and Wenbing Huang and Peijia Lin and Jiaqi Han and Pin Chen and Yutong Lu and Yang Liu},
  title = {{Crystal Structure Prediction by Joint Equivariant Diffusion}},
  booktitle = {Advances in Neural Information Processing Systems (NeurIPS)},
  year = {2023},
  url = {https://arxiv.org/abs/2309.04475}
}

@inproceedings{Jiao2024DiffCSPpp,
  author = {Rui Jiao and Wenbing Huang and Yu Liu and Deli Zhao and Yang Liu},
  title = {{Space Group Constrained Crystal Generation}},
  booktitle = {International Conference on Learning Representations (ICLR)},
  year = {2024},
  url = {https://arxiv.org/abs/2402.03992}
}

@inproceedings{Miller2024FlowMM,
  author = {Benjamin Kurt Miller and Ricky T. Q. Chen and Anuroop Sriram and Brandon M. Wood},
  title = {{FlowMM: Generating Materials with Riemannian Flow Matching}},
  booktitle = {Proceedings of the 41st International Conference on Machine Learning (ICML)},
  year = {2024},
  url = {https://arxiv.org/abs/2406.04713}
}

@misc{Widdowson2021PDD,
  author = {Daniel Widdowson and Vitaliy Kurlin},
  title = {{Pointwise Distance Distributions for detecting near-duplicates in large materials databases}},
  howpublished = {arXiv preprint arXiv:2108.04798},
  year = {2021}
}

@misc{Wei2026GenerativeCSPBenchmark,
  author = {Lai Wei and Rongzhi Dong and Ying Feng and Madeline Miklos and Jianjun Hu},
  title = {{Deep Generative Crystal Structure Prediction: A Benchmark Study and a Controlled Test of Prototype Dependence}},
  howpublished = {arXiv preprint arXiv:2609.26502},
  year = {2026}
}

@inproceedings{Lin2024EquiCSP,
  author = {Peijia Lin and Pin Chen and Rui Jiao and Qing Mo and Jianhuan Cen and Wenbing Huang and Yang Liu and Dan Huang and Yutong Lu},
  title = {{Equivariant Diffusion for Crystal Structure Prediction}},
  booktitle = {Proceedings of the 41st International Conference on Machine Learning (ICML)},
  year = {2024},
  url = {https://arxiv.org/abs/2512.07289}
}

@misc{Lu2025Uni3DAR,
  author = {Shuqi Lu and Haowei Lin and Lin Yao and Zhifeng Gao and Xiaohong Ji and Yitao Liang and Weinan E and Linfeng Zhang and Guolin Ke},
  title = {{Unified Cross-Scale 3D Generation and Understanding via Autoregressive Modeling}},
  howpublished = {arXiv preprint arXiv:2503.16278},
  year = {2025}
}

@article{Wei2024CCT,
  author = {Lai Wei and Qinyang Li and Yuqi Song and Stanislav Stefanov and Rongzhi Dong and Nihang Fu and Edirisuriya M. D. Siriwardane and Fanglin Chen and Jianjun Hu},
  title = {{Crystal Composition Transformer: Self-Learning Neural Language Model for Generative and Tinkering Design of Materials}},
  journal = {Advanced Science},
  volume = {11},
  pages = {2304305},
  year = {2024},
  doi = {10.1002/advs.202304305}
}

@article{Fu2023OxidationState,
  author = {Nihang Fu and Jeffrey Hu and Ying Feng and Gregory Morrison and Hans-Conrad zur Loye and Jianjun Hu},
  title = {{Composition Based Oxidation State Prediction of Materials Using Deep Learning Language Models}},
  journal = {Advanced Science},
  volume = {10},
  pages = {2301011},
  year = {2023},
  doi = {10.1002/advs.202301011}
}

@article{Fu2023MaterialTransformers,
  author = {Nihang Fu and Lai Wei and Yuqi Song and Qinyang Li and Rui Xin and Sadman Sadeed Omee and Rongzhi Dong and Edirisuriya M. Dilanga Siriwardane and Jianjun Hu},
  title = {{Material Transformers: Deep Learning Language Models for Generative Materials Design}},
  journal = {Machine Learning: Science and Technology},
  volume = {4},
  number = {1},
  pages = {015001},
  year = {2023},
  publisher = {IOP Publishing}
}

@article{Hu2024GenerativeAI,
  author = {Jianjun Hu and Qin Li and Nihang Fu},
  title = {{Generative AI for Materials Discovery: Design Without Understanding}},
  journal = {Engineering},
  volume = {39},
  pages = {13--17},
  year = {2024},
  doi = {10.1016/j.eng.2024.07.008}
}

@article{Luo2025CrystalFlow,
  author = {Xiaoshan Luo and Zhenyu Wang and Qingchang Wang and Xuechen Shao and Jian Lv and Lei Wang and Yanchao Wang and Yanming Ma},
  title = {{CrystalFlow}: A Flow-Based Generative Model for Crystalline Materials},
  journal = {Nature Communications},
  volume = {16},
  number = {1},
  pages = {9267},
  year = {2025}
}

@article{Dey2026uFlowCSP,
  author = {Sourin Dey and Dipannoy Das Gupta and Lai Wei and Sadman Sadeed Omee and Jianjun Hu},
  title = {{uFlowCSP}: Crystal Structure Prediction using Mean Flow Generative Models},
  journal = {arXiv preprint arXiv:2609.09799},
  year = {2026}
}

@article{Zhao2021CubicGAN,
  author = {Yong Zhao and Mohammed Al-Fahdi and Ming Hu and Edirisuriya M. Dilanga Siriwardane and Yuqi Song and Alireza Nasiri and Jianjun Hu},
  title = {High-Throughput Discovery of Novel Cubic Crystal Materials Using Deep Generative Neural Networks},
  journal = {Advanced Science},
  volume = {8},
  number = {20},
  pages = {2100566},
  year = {2021}
}

@article{Zhao2023PGCGM,
  author = {Yong Zhao and Edirisuriya M. Dilanga Siriwardane and Z. Wu and Nihang Fu and Mohammed Al-Fahdi and Ming Hu and Jianjun Hu},
  title = {Physics Guided Deep Learning for Generative Design of Crystal Materials with Symmetry Constraints},
  journal = {npj Computational Materials},
  volume = {9},
  number = {1},
  pages = {38},
  year = {2023}
}
\end{document}